%% file: main.tex
\documentclass[conference,hyphens]{IEEEtran}
\usepackage[utf8]{inputenc}
\usepackage[linesnumbered, ruled]{algorithm2e}
\usepackage[noend]{algpseudocode}

\usepackage{cite}
\usepackage{amsmath,amssymb,amsfonts}
\usepackage{graphicx}
\usepackage{textcomp}
\usepackage{xcolor}
\usepackage{booktabs}
\usepackage{tagging}
\usepackage{ipdps26repro}
\usepackage[hidelinks]{hyperref} 
\hypersetup{ 
    colorlinks=true,
    linkcolor=blue,
    filecolor=blue,      
    urlcolor=blue,
}
\usepackage{array, makecell}
\usepackage[inline]{enumitem}
\usepackage{cite}
\usepackage{caption}
\usepackage{amsmath,amssymb,amsfonts}
\usepackage{float}
\usepackage[caption=false, font=footnotesize]{subfig}
\usepackage[font={small}]{caption}
\usepackage{booktabs}

\usepackage{tikz}
\usepackage{pgfplots,filecontents}
\usepackage{cprotect}
\usepackage{wrapfig}

\usepackage{mathtools}
\usepackage{xpatch}
\usepackage{xcolor}
\usepackage{realboxes}
\usepackage{verbatim}
\usepackage{bm}
\usepackage{multirow}
\usepackage{listings}
\definecolor{codegreen}{rgb}{0,0.6,0}
\definecolor{codegray}{rgb}{0.5,0.5,0.5}
\definecolor{codepurple}{rgb}{0.58,0,0.82}
\definecolor{backcolour}{rgb}{0.95,0.95,0.92}
\lstdefinestyle{terminal}{
  basicstyle=\ttfamily\footnotesize,
  backgroundcolor=\color{backcolour},
  commentstyle=\color{codegreen},
  keywordstyle=\color{magenta},
  numberstyle=\tiny\color{codegray},
  stringstyle=\color{codepurple},
  breaklines=true,
  postbreak=\mbox{\textcolor{red}{$\hookrightarrow$}\space},
  showstringspaces=false,
  keywordstyle=\color{blue},
  commentstyle=\color{green},
  stringstyle=\color{red}
}
\usepackage{bbding} 
\lstdefinestyle{sqlstyle}{
    language=SQL,
    basicstyle=\ttfamily\small,
    keywordstyle=\color{blue}\bfseries,
    stringstyle=\color{teal},
    commentstyle=\color{gray},
    frame=single,
    framerule=0.1pt,
    rulecolor=\color{black},
    backgroundcolor=\color{backcolour},
    showstringspaces=false,
    breaklines=true,
}

\usepackage{graphicx}
\usepackage[binary-units]{siunitx}
\DeclareSIUnit\event{event}
\usepackage{textcomp}
\usepackage{xcolor}
\usepackage{soul}
\usepackage{orcidlink}

\usepackage[caption=false]{subfig}
\def\BibTeX{{\rm B\kern-.05em{\sc i\kern-.025em b}\kern-.08em
    T\kern-.1667em\lower.7ex\hbox{E}\kern-.125emX}}

\newcommand{\upp}{\vspace*{-0.5em}}

\begin{document}

\title{
    Beyond Thread States: Diagnosing Performance Degradation with eBPF and Thread Dynamics
    %\upp
}

\author{\IEEEauthorblockN{Diogo Landau\orcidlink{0009-0000-2625-8884
}\IEEEauthorrefmark{1},
Jorge G. Barbosa\orcidlink{0000-0003-4135-2347}\IEEEauthorrefmark{2}, Nishant Saurabh\orcidlink{0000-0002-1926-4693}\IEEEauthorrefmark{1}} 
\IEEEauthorblockA{\IEEEauthorrefmark{1}\textit{Department of Information and Computing Sciences, Utrecht University, NL}}
\IEEEauthorrefmark{2}{\textit{LIACC, Faculdade de Engenharia, Universidade do Porto, PT}}\\
d.landau@uu.nl\IEEEauthorrefmark{1},
jbarbosa@fe.up.pt\IEEEauthorrefmark{2},
n.saurabh@uu.nl\IEEEauthorrefmark{1}
%\upp\upp\upp
}

\maketitle

\begin{abstract}
Online Data-Intensive applications face performance degradation from load variability and resource interference. While Thread State Analysis (TSA) based approaches enable identifying constrained subsystems, they lack the granularity to reveal the inter-thread dependencies that propagate degradation. In this paper, we present an application-agnostic performance degradation analysis method that extends TSA by capturing fine-grained thread dynamics. We implemented $16$ \textit{eBPF-based metrics} across six kernel subsystems, including scheduling, VFS, networking, futex, multiplexing IO, and block IO which enables tracing thread interactions with specific resources like futexes, sockets, and disks. Our method leverages the fact that performance degradation propagates along inter-thread dependencies, and a subset of thread-resource interactions can enable capturing common degradation patterns. To this end, we employ a selective thread tracking algorithm that traces performance issues from entry-point threads to constrained resources. Experimentation with diverse applications under variable workloads and resource contention shows our method successfully diagnoses CPU, disk, lock, and external service contention with minimal overhead, while also revealing internal application constraints.

\end{abstract}


\maketitle

%-------------------------------------------------------------------------------
\section{Introduction}
%-------------------------------------------------------------------------------
%\textit{Motivation:} 
The modern internet is often powered by a vast network of  interconnected applications (e.g., online data-intensive applications ~\cite{meisner2011power, sriraman2018mutune} such as databases, message brokers, and machine learning solutions), providing data, services, and functionality to connecting processes ~\cite{sriraman2018mutune}. Load variability and interference can degrade these applications' Quality of Service (QoS), negatively affecting dependent applications and end-user experience~~\cite{manghwani2005end,li2020amoeba, arapakis2014impact, bouch2000quality}. Restoring QoS requires diagnosing the cause of degradation to apply suitable mitigation ~\cite{dean2014perfcompass}. This necessitates a solution capable of \textit{observing} an application's \textit{performance} by collecting appropriate metrics to understand its state.

%\textit{Limitations of state-of-art approaches:} 
An established approach to performance investigation is Thread State Analysis (TSA)~\cite{gregg2014systems}, which quantifies the time each application thread spends in six distinct states (Executing, Runnable, Swapping, Sleeping, Lock, and Idle). TSA provides valuable insight into the subsystems where application threads spend time and helps identify performance bottlenecks within those subsystems (e.g., CPU contention, I/O wait, or synchronization delays). However, TSA alone cannot reveal the interactions among threads that collectively characterize an application’s performance. This limitation arises because TSA collects metrics at a per-thread granularity, making it opaque to the specific system resources accessed by threads and, consequently, to the thread dynamics that emerge from these interactions. These \textit{thread dynamics}~\cite{sites2021understanding} are essential for uncovering common performance degradation patterns related to specific system resources. 

\begin{figure}[hbt!]
    \centering
   \upp\upp\upp\upp
    \subfloat[\label{fig:database:latency}]{
      \includegraphics[width=0.35\linewidth]{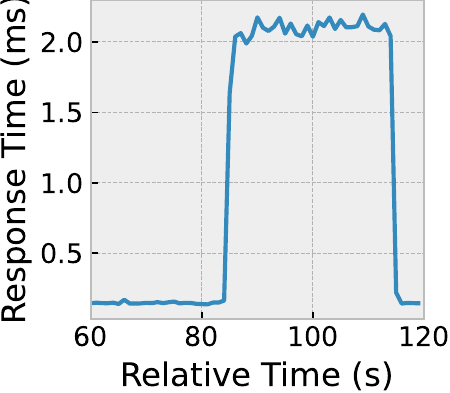} 
    }
    \subfloat[\label{fig:database:lock_wait}]{
      \includegraphics[width=0.35\linewidth]{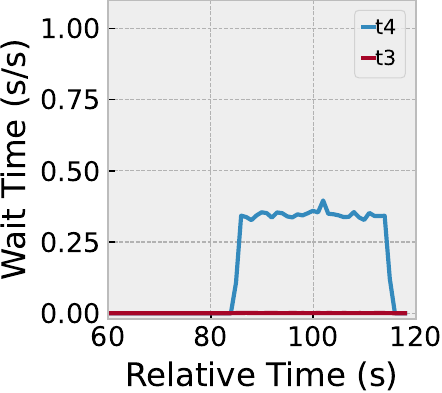} 
    }
    \subfloat[\label{fig:database:graph}]{
      \includegraphics[height=0.35\linewidth]{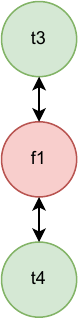} 
    }
    %\upp 
    \caption{\textbf{Illustration of a database experiencing lock contention, leading to degraded performance:} (a) $95^{th}$ percentile response time; (b) Total time database threads spend waiting for locks; (c) Visualization of database thread dynamics between revealing threads \texttt{t3} and \texttt{t4} interacting via a lock \texttt{f1}.}
   %\upp
\end{figure}

Figure~\ref{fig:database:latency} illustrates this through a database experiencing degraded performance caused by the serialization of read and write operations on a shared row. TSA results in Figure~\ref{fig:database:lock_wait} indicate that the increased time spent by thread \texttt{t4} waiting on locks strongly correlates with the observed performance degradation. However, the thread dynamics graph in Figure~\ref{fig:database:graph} further reveals that the only other thread interacting with \texttt{t4} is \texttt{t3}, through a shared lock. This interaction highlights that the degraded performance originates from thread \texttt{t3}, which is processing a sudden surge of write operations.

Traditionally, performance analysis approaches have relied on system-level metrics exposed by the operating system (e.g., the Linux \texttt{/proc} filesystem) to infer resource constraints and performance bottlenecks~\cite{grohmann2019monitorless, cheng2024fedge, yanggratoke2015predicting, ahmed2018automated, cohen2004correlating, zhang2005ensembles, gan2021sage, gan2019seer}. While such metrics provide valuable insights, they often lack the granularity needed to capture an application’s dynamics. Recent advances in the Linux kernel, particularly the development of the extended Berkeley Packet Filter (eBPF) subsystem, offer new opportunities to overcome these limitations. Building on eBPF, emerging approaches \cite{rezvani2024characterizing, jha2022holistic} have begun to collect detailed thread state metrics that provide richer context for diagnosing performance degradation. 

However, these methods still fall short in capturing the fine-grained interactions necessary to fully uncover an application’s thread dynamics. A notable effort in this direction is presented by~\cite{seo2022nosql}, which defines a set of metrics for identifying lock-related dynamics in NoSQL applications. While promising, this approach remains domain-specific and focuses solely on sychronization behaviour. In contrast, our goal is to generalize this capability to encompass a broader range of system resources and application types.

%\textit{Contributions and Methodology:} 
This paper introduces a set of TSA metrics that, through carefully tuned granularity, also expose the thread dynamics that lead to an application's experienced degraded performance. To achieve this, we implement $16$ \textit{eBPF-based metrics} across six kernel subsystems, including scheduling, virtual file system (VFS), networking, futex, multiplexing IO, and block IO. These metrics enhance observability by tracking a thread's use of specific kernel resources, revealing common degradation scenarios like lock, disk, CPU, and external service dependency contention. Furthermore, we employ selective thread tracking to analyse application degradation under the lens of its thread dynamics, to reveal how request-facing threads are contended by resource-constrained threads within an application.

To validate our approach, we evaluated our method on an extensive set of online data-intensive applications, offering variability in application domains, threading models, inter-process communication (IPC) methods, and user-space virtual machines. Our experiments considers scenarios where applications degrade under a combination of variable workload patterns and resource types. 

Our results show that the collected metrics successfully highlight the constrained resources and the thread dynamics that led to an application's degraded state. In the presence of lock contention scenarios, the metrics enabled identifying the threads and futex resources hindering application response time. For disk contention, they revealed the application's inter-thread communication model, linking server response time to disk performance. In CPU-constrained cases, the implemented metrics highlighted impaired threads, identifying runqueue activity as the primary cause of degraded performance. Finally, when degradation originated from an external service dependency, metrics revealed the external service impairing performance, indicating the issue was outside the application under analysis. We have so far analyzed how performance degradation manifests at the thread level; however, we aim to extend diagnosability by integrating system-level information to obtain a more comprehensive view of an application's performance, which we plan to address in our future work. The complete source code and reproducibility artifact is released opensource in an \emph{online} \verb|GitHub| repository~\cite{repo}. 

%-------------------------------------------------------------------------------
\section{Related Work}\label{sec:related}
%-------------------------------------------------------------------------------
This section discusses state-of-the-art system instrumentation methods employed in performance degradation analysis related works.

%\subsection{System-level Instrumentation}
%\paragraph{\textbf{System-level Instrumentation}} 
\textbf{System-level Instrumentation:} Application-agnostic performance monitoring methods collect system-level metrics to understand the application's interaction with the kernel, virtualized resources or the underlying hardware. One such method is sampling, which collects system-level metrics at periodic intervals to derive statistics encoding an application's activity~\cite{nagar2007your, gavin1998performance}. Another method is binary instrumentation~\cite{kreutzer2023runtime, seo2022nosql, wang2024diagnosing}, which dynamically inserts probes in the application to collect custom performance statistics. 

\emph{SystemTap}~\cite{prasad2005locating}, \emph{LTTng}~\cite{desnoyers2006lttng} and \emph{Dtrace}~\cite{cantrill2004dynamic} are early solutions which dynamically instrument kernel-level code through dedicated kernel modules. However, their unrestricted kernel memory access frequently triggered kernel panics, rendering them unsuitable for production environments. Nevertheless, \emph{Extended Berkeley Packet Filter (eBPF)}~\cite{ebpf} addressed this issue by implementing and utilizing a kernel virtual machine which restricts \emph{ebpf} program's memory access through specialised helper functions, while interpreting probe instructions. 

Typically, \emph{eBPF} probes are event-driven lightweight programs that are placed by specific system calls into kernel memory, and executed in a protected environment. Such programs are often difficult to develop, primarily due to its complex interactions with kernel data structures, events and its sandboxed execution. However, tools like \emph{BCC}~\cite{bcc}, \emph{bpftrace}~\cite{bpftrace}, \emph{libbpf}~\cite{libbpf} \emph{libbpfrs}~\cite{libbpfrs} abstract \emph{eBPF's} low-level details and provide a high-level interface to develop \emph{ebpf} programs. In this paper, we use \emph{libbpf}~\cite{bpftrace} to strategically trace particular kernel functions (see details in Section~\ref{sec:methodology}).

\textbf{Performance Degradation Analysis:} 
Root-cause analysis (RCA) is central to diagnosing application performance degradation~\cite{wyatt2020canario}. Statistical approaches such as BARO~\cite{pham2024baro} and N-Sigma~\cite{li2022causal,lin2018microscope} detect distribution shifts in time-series metrics before and after anomalies to flag potential causes. Similarly, prior work~\cite{cohen2004correlating, zhang2005ensembles, gan2021sage, gan2019seer} correlates hardware resource usage with performance anomalies, while systems like FIRM~\cite{qiu2020firm} apply learning-based models to infer metric relevance for performance variability within microservice dependency graphs. While the use of readily available hardware metrics makes these methods broadly applicable, it also limits diagnostic depth, omitting the fine-grained behavioral context necessary to accurately identify the causes of application-level performance degradation.

With the emergence of \emph{eBPF} subsystem, recent work has expanded observability beyond hardware resources to include detailed kernel-level interactions. Rezvani et al.~\cite{rezvani2024characterizing} leverage \emph{eBPF} to analyze the relationship between application load and the \emph{epoll} subsystem, while Jha et al.~\cite{jha2022holistic} instrument the virtual file system and block I/O layers to detect anomalous application behavior. Other solutions, such as Coroot~\cite{coroot} and Pixie~\cite{pixie}, employ \emph{eBPF} to capture process-level metrics related to service communication and hardware resource usage. 
{\color{black}
In contrast, APM solutions provided by Dynatrace~\cite{dynatraceAPM} and New Relic~\cite{newrelicAPM} enhance application observability through language-specific agents that utilize bytecode instrumentation to capture transaction-level context and runtime specific metrics; however, they are limited to userspace behavior and do not observe thread interactions with kernel resources. 
% In contrast, \emph{non-eBPF} based commercial APM tools by Dynatrace~\cite{dynatraceAPM}, New Relic~\cite{newrelicAPM} enhance application observability through language-specific agents that utilize bytecode instrumentation to capture transaction-level context and trace requests through application code, methods and database queries. 
% Commercial APM solutions provided by Dynatrace~\cite{dynatraceAPM} and New Relic~\cite{newrelicAPM} enhance application observability through language-specific instrumentation integrated into applications via dedicated libraries. Our work is complementary, providing additional visibility into thread-level interactions with system resources that can be correlated with application-level metrics during performance degradation analysis.
}
Consequently, these tools remain opaque to important application \emph{thread dynamics}, that reveal interactions between threads mediated by kernel resources such as locks, sockets, pipes, and other inter-process communication mechanisms, which are essential for diagnosing the propagation paths of performance degradation within a host.

\begin{figure}[t]
    \centering
    \includegraphics[width=\linewidth]{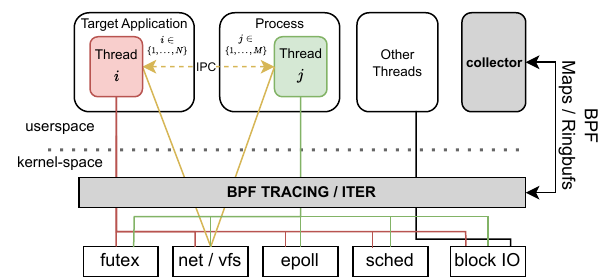}
    \caption{{\textbf{Instrumentation architecture:} 
    Illustration of the required components to monitor a target application. Our instrumentation also accounts for IPC and therefore thread groups that have communicated with the target application will also be monitored.}} 
    \upp\upp
    \label{fig:architecture}
\end{figure}

\begin{figure*}[t]
    \centering
    \includegraphics[width=\linewidth]{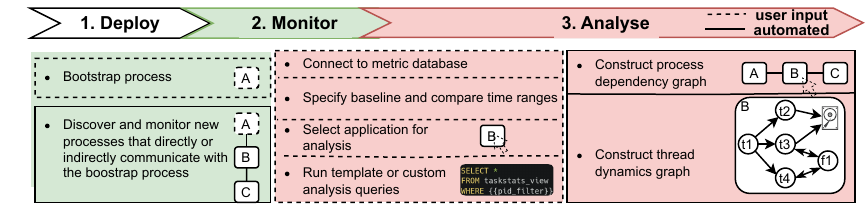}
    \caption{{\textbf{Diagnostic Workflow:} 
     overview of the diagnostic workflow for application performance degradation. Stages highlighted in red and green denote our method’s participation and include a summary of the associated operations, indicating whether they are automated or require user intervention.}} 
    %\upp\upp\upp
    \label{fig:diagnostic-workflow}
\end{figure*}

% \begin{figure*}[hbt!]
%     \centering
%     \upp\upp\upp\upp
%     \subfloat[\label{fig:architecture}]{
%       \includegraphics[width=0.35\linewidth]{figs/arch-new.pdf} 
%     }
%     \subfloat[\label{fig:diagnostic-workflow}]{
%       \includegraphics[width=0.63\linewidth]{figs/Diagnostic-Workflow-revised.pdf} 
%     }
%     \upp 
%     \caption{{(a) \textbf{Instrumentation architecture:} 
%     Illustration of the required components to monitor a target application. Our instrumentation also accounts for IPC and therefore thread groups that have communicated with the target application will also be monitored.}  \color{blue} (b) \textbf{Diagnostic Workflow:} 
%     overview of the diagnostic workflow for application performance degradation. Stages highlighted in red and green denote our method’s participation and include a summary of the associated operations, indicating whether they are automated or require user intervention.
%     }
%    \upp
% \end{figure*} 

The importance of capturing such thread dynamics is underscored by~\cite{sriraman2018mutune,sites2021understanding}, who demonstrate that an application’s threading model plays a decisive role in its performance. 
{\color{black}
KUTrace\cite{sites2021understanding} exposes these dynamics and achieves full-system coverage by instrumenting kernel-to-userspace transitions. However, its kernel-boundary observation cannot identify specific coordination resources (e.g., locks) without injecting custom userspace  C libraries, introducing both build-time and runtime dependencies. While prioritizing low overhead and full state coverage, it generates 2-20~MB/s of trace data to RAM, (upto 2.4~GB over 120~s). Our approach eliminates these dependencies and reduces data volume, enabling longer-term monitoring while preserving observability.

% KUTrace\cite{sites2021understanding} exposes these dynamics by instrumenting all kernel-to-userspace transitions, achieving full-system coverage. However, because it only observes transitions at the kernel  boundary, it cannot identify the specific system resources (e.g., locks) involved in thread coordination without injecting additional userspace context using custom C libraries. This introduces both build-time and runtime dependencies. Moreover, KUTrace prioritizes low runtime overhead and full thread-state coverage, generating 2-20~MB/s trace data to RAM, accumulating upto 2.4~GB over 120~s. Our approach aims to address these concerns by eliminating instrumentation dependencies and reducing data volume, enabling longer-term monitoring without compromising observability.
}

% {\color{blue}
% KUTrace \cite{sites2021understanding} is explicitly designed to expose application thread dynamics and therefore instruments all kernel to userspace transitions to achieve full-system coverage. However, because its observation point is limited to the kernel–userspace boundary, the context available at this level is insufficient to fully determine the specific system resources (e.g., locks) that coordinate thread interactions. To address this limitation, KUTrace provides a userspace C library that injects additional contextual information at runtime, enabling the identification of locking primitives involved in thread synchronization. As a consequence, achieving this level of observability, requires bundling the software with KUTrace's complementary libraries. Furthermore, given that it prioritizes thread state coverage and low runtime processing overhead, KUTrace produces a substantial volume of trace data, producing approximately 2-20~MB/s to memory, which accumulates upto 2.4~GB over roughly 120~s of execution. We aim to address both concerns by elimiinating build- and run-time dependencies when constructing a process’s thread dynamics, while also reducing the volume of data produced during instrumentation to enable monitoring an application over longer durations.
% }

To provide more granular observability, xCapture~\cite{xCapture} measures the time each application thread spends within Linux kernel subsystems, enabling Thread State Analysis (TSA). However, its metrics lack resource-level context and therefore cannot capture the inter-thread interactions that underlie application performance behavior. Seo et al.~\cite{seo2022nosql} address this limitation for NoSQL databases by analyzing thread dynamics based on futex interactions. Building on this direction, our work generalizes this capability to a wider range of kernel resources and application types.

% Building on this direction, our work generalizes this capability to a wider range of kernel resources and application types. Specifically, to enable TSA, our system collects metrics at a thread and kernel resource granularity, capturing both the time each thread spends in key kernel subsystems and the interactions through which threads coordinate. This added dimension  enables the discovery of performance degradation patterns driven by thread dynamics that remain opaque to prior approaches.

{\color{black}\textbf{Comparison Summary}: 
Dynatrace~\cite{dynatraceAPM} and NewRelic~\cite{newrelicAPM} provide application-layer observability, but require bytecode-agent instrumentation, which modifies runtime execution and can incur high overhead in terms of increased application response time, memory and CPU usage~\cite{newrelicReduceInfrastructure,newrelicTroubleshootingLarge, dynatraceControllingMeasurement,dynatraceProcessDeep} in default mode, compared to kernel instrumentation. Coroot~\cite{coroot} and Pixie~\cite{pixie} are application-agnostic but operate at service/process granularity. KUTrace~\cite{sites2021understanding} and xCapture~\cite{xCapture} provide thread-level tracing at finer granularity, but lack resource-level context needed to model inter-thread dependencies. KUTrace, however, provides custom instrumentation libraries that can be integrated into application software to expose additional context on lock usage. Since these libraries must be incorporated at build-/run- time, thread dynamics remain opaque for third-party applications that do not use them.

% Therefore, a generalized application-agnostic method is needed that extends TSA with additional kernel resource-thread context to expose degradation propagation with low overhead. Our approach addresses this by implementing targeted eBPF probes across six kernel subsystems, by collecting metrics at \textit{``thread$\rightarrow$resource''} granularity, our approach tracks the time a thread spends in a particular subsystem, but also tracks which resources within this subsystem it is spending time on. Accordingly, our approach complements state-of-the-art observability solutions rather than replacing them, as they continue to provide essential performance analysis data, including distributed tracing and profiling.

Therefore, a generalized, application-agnostic method is needed to extend TSA with additional kernel-level thread–resource context in order to expose degradation propagation with low overhead. Our approach addresses this need by implementing targeted eBPF probes across six kernel subsystems. By collecting metrics at \textit{``thread$\rightarrow$resource''} granularity, it captures both the time a thread spends within a given subsystem and the specific resources on which this time is spent. Accordingly, this approach complements state-of-the-art observability solutions rather than replacing them, as they continue to provide essential performance analysis capabilities, including distributed tracing and profiling.

% Therefore, a generalized application-agnostic method is needed that performs TSA at the granularity of specific kernel resources to expose degradation propagation with low overhead. Our approach addresses this by implementing targeted eBPF probes across six kernel subsystems, enabling TSA by collecting metrics at \textit{``thread$\rightarrow$resource''} granularity that capture both per-thread time in key kernel subsystems and fine-grained interactions through which threads coordinate. This enables the discovery of performance degradation patterns rooted in thread dynamics opaque to SOTA tools, but complement rather than replacing them by adding system-level insight into thread-resource interactions.
}

\section{Design and Architecture}\label{sec:archi}
%-------------------------------------------------------------------------------
Thread State Analysis (TSA) requires tracking the time each application thread spends in six distinct states~\cite{gregg2014systems}: Executing, Runnable, Swapping, Sleeping, Lock, and Idle. Some of these states can be further subdivided~\cite{greggTSAblog} to provide finer-grained insight into where threads spend their time. For example, Sleeping time may be partitioned into more specific categories such as waiting on storage, network, or pipe operations.

However, our goal extends beyond simply measuring how long threads spend in these subsystems. We also aim to capture how threads interact with one another through shared system-level resources. These inter-thread dependencies often reveal how performance degradation originating from a constrained resource propagates across threads, thereby exposing the application’s critical path, i.e., the sequence of dependent threads and resources that ultimately drives performance degradation.

To this end, given the defined set of \textit{Thread States}, we identify and collect a comprehensive list of metrics required to perform TSA, as shown in Table~\ref{table:metrics}. We exclude the \emph{Swapping} state from collection, since it typically arises under system-wide memory pressure rather than from thread-specific activity. Nevertheless, each metric corresponds to one of the six states and, where applicable, includes finer-grained subdivisions based on its \textit{Subsystem}. To reveal an application’s thread dynamics, these metrics are collected at a resource-level granularity, enabling the attribution of thread time to specific system-level resources. For instance, \textit{``thread$\rightarrow$futex''} \textit{Granularity} indicates, \emph{futex\_wait\_\{time,count\}} metrics are collected per thread for a specific futex resource. This approach not only preserves the traditional TSA view of per-thread state distribution but also exposes inter-thread dependencies by capturing how threads wait on and interact through shared resources, often used as synchronization primitives. Table~\ref{table:metrics} summarizes the full set of collected metrics and corresponding thread states.

\begin{table*}[!t]
\upp
\centering
    \caption{Enumeration of the instrumented metrics based on the Thread States enumerated by TSA~\cite{gregg2014systems}, and the resource granularity required to monitor thread dynamics.}
\label{table:metrics}
\upp
%\vspace*{-0.1cm}
\setlength{\tabcolsep}{0.3pt}
\resizebox{0.7\linewidth}{!}
{
\begin{tabular}{|c|c|c|c|c|}
\cline{1-5}
%\hline
\hline
    \textbf{State}&\textbf{Subsystem} & \textbf{Metric} & \textbf{Granularity} & \textbf{Metric description}
\\
\cline{1-5}
\hline
\hline

        Executing & \multirow{5}{*}{Scheduler} & \emph{runtime} & thread & time spent on-CPU \\
        \cline{1-1}\cline{3-5}
        Runnable & & \emph{rq\_time} & thread & time waiting on runqueue \\
        \cline{1-1}\cline{3-5}
        \multirow{3}{*}{Sleeping/} & & \emph{block\_time} & thread & time in uninterruptible sleep \\
        \cline{3-5}
        \multirow{3}{*}{Idle}& & \emph{iowait\_time} & thread & time in uninterruptible sleep and in \emph{iowait}  \\
        \cline{3-5}
        & & \emph{sleep\_time} & thread & time spent in interruptible sleep \\
        \cline{3-5}
        \hline
        \hline
        
        \multirow{5}{*}{Sleeping/} & \multirow{5}{*}{\shortstack{VFS,\\ Network, IO\\ Multiplexing,\\Block IO}} & \emph{pipe\_wait\_time} & thread$\rightarrow$pipe &  time thread waits for pipe \\
        \cline{3-5}
        \multirow{5}{*}{Idle}& & \emph{pipe\_wait\_count} & thread$\rightarrow$pipe & thread pipe wait frequency \\
        \cline{3-5}
        & & \emph{socket\_wait\_time} & thread$\rightarrow$socket & time thread waits for socket \\
        \cline{3-5}
        & &\emph{socket\_wait\_count} & thread$\rightarrow$socket & thread socket wait frequency \\
        \cline{3-5}
        & & \emph{sector\_count} & thread$\rightarrow$device & thread sector requests to device \\
        \cline{3-5}
       \hline
       \hline
       \multirow{3}{*}{Sleeping/}  & \multirow{3}{*}{\shortstack{Multiplexing}} & \emph{epoll\_wait\_time} & thread$\rightarrow$epoll & thread epoll wait time \\
       \cline{3-5}
        \multirow{3}{*}{Idle}& \multirow{3}{*}{\shortstack{IO}} & \emph{epoll\_wait\_count} & thread$\rightarrow$epoll & thread epoll wait frequency \\
        \cline{3-5}
        & & \emph{epoll\_file\_wait} & epoll$\rightarrow$virtual file & time epoll waits for file \\
        \cline{3-5}
        \hline
        \hline
        \multirow{3}{*}{Lock} & \multirow{3}{*}{\shortstack{Futex}} & \emph{futex\_wait\_time} & thread$\rightarrow$futex & time waiting for a futex \\
        \cline{3-5}
        & & \emph{futex\_wait\_count} & thread$\rightarrow$futex & futex wait frequency \\
        \cline{3-5}
        & & \emph{futex\_wake\_count} & thread$\rightarrow$futex & futex wake frequency \\
        \cline{3-5}
        \hline
\end{tabular} 
}
%\upp\upp
\end{table*}

% \begin{figure}[t]
%     \centering
%     \includegraphics[width=0.65\linewidth]{figs/arch-new.pdf}
%     \caption{{\textbf{Instrumentation architecture:} 
%     Illustration of the required components to monitor a target application. Our instrumentation also accounts for IPC and therefore thread groups that have communicated with the target application will also be monitored.}} 
%     \upp\upp
%     \label{fig:architecture}
% \end{figure}

% \subsection{Instrumentation Architecture}
% \paragraph{\textbf{Instrumentation Architecture}} 
\textbf{Instrumentation Architecture}: Figure~\ref{fig:architecture} depicts our instrumentation architecture. When the target application executes probed kernel functions, custom eBPF programs capture raw statistics on thread interactions with the six kernel subsystems. These statistics are communicated to a userspace component, the collector, via maps and ring buffers, which then processes the raw data into the metrics summarized in Table~\ref{table:metrics}. To reduce resource overhead, the collector reads only statistical summaries at one-second intervals, even though the eBPF programs trace events at a much higher frequency. In the following Section, we describe the probes and techniques employed to realize the metrics described in Table~\ref{table:metrics}.

{\color{black}

% \begin{figure}[t]
%     \centering
%     \includegraphics[width=\columnwidth]{figs/Diagnostic-Workflow-revised.pdf}
%     \caption{
%     {\color{blue}\textbf{Diagnostic Workflow} 
%     High-level overview of the diagnostic workflow for application performance degradation. Stages highlighted in red and green denote our method’s participation and include a summary of the associated operations, indicating whether they are automated or require user intervention.
%     }}
%     \upp\upp
%     \label{fig:diagnostic-workflow}
% \end{figure}

\textbf{Diagnostic Workflow}: 
Figure~\ref{fig:diagnostic-workflow} illustrates the high-level diagnostic workflow of our approach with three stages.  

% \emph{(a)} The \emph{application deployment}, is external to our method. This eliminates the need for recompilation, modification or prior instrumentation.
\emph{(a)} The \emph{application deployment}, is external to our method. Excluding application deployment as a requirement, is a key design objective which further eliminates the need for recompilation, modification or prior instrumentation.
Consequently, our approach supports a wide range of applications and enables runtime monitoring on any sufficiently recent Linux system with eBPF support.
% % , compared to tools like KUTrace\cite{sites2021understanding} with build and runtime dependencies. 
% Consequently, our approach supports a wide range of applications and enables runtime monitoring on any sufficiently recent Linux system with eBPF support.

\emph{(b)} The \emph{monitoring stage} initiates when a user specifies a bootstrapping process (e.g., a service’s main process) for the metric collector. The system then automatically deploys eBPF probes to track the specified process and dynamically discovers all directly and indirectly interacting processes through IPC mechanisms such as sockets and pipes. This discovery is transitive, constructing a complete process dependency graph without manual intervention (e.g., in Figure~\ref{fig:diagnostic-workflow}, if process~A communicates with B, and B with C, all are monitored and automatically discovered). The set of instrumented metrics (Table~\ref{table:metrics}) are collected for all discovered processes and stored in an OLAP database \cite{raasveldt2019duckdb}.
% \emph{(b)} The \emph{monitoring stage} initiates when a user specifies a bootstrapping process (e.g., a service’s main process) for the metric collector. Our system automatically employs eBPF probes and begins tracking it. It then dynamically discovers all (in)directly interacting processes through IPC mechanisms (e.g. sockets and pipes). This discovery is transitive, constructing a complete process dependency graph without manual intervention (e.g., in Figure~\ref{fig:diagnostic-workflow}, if process~A communicates with B, and B with C, all are monitored and automatically discovered). The set of instrumented metrics (Table~\ref{table:metrics}) are collected for all discovered processes and stored in an OLAP database \cite{raasveldt2019duckdb}.
% \emph{(b)} The \emph{monitoring stage} initiates when a user specifies a  bootstrapping process (e.g., a service’s main process). Our system automatically deploys eBPF probes and begins tracking the specified process. It then dynamically discovers all (in)directly interacting processes through inter-process communication (IPC) mechanisms (e.g. sockets and pipes). This discovery is transitive: for example, in Figure~\ref{fig:diagnostic-workflow}, process~A communicates with B, and B with C, all are monitored, and automatically discovered, constructing a complete process dependency graph without manual intervention. The set of instrumented metrics (Table~\ref{table:metrics}) is collected for all discovered processes, as detailed in Section~\ref{sec:collection} and are stored in an OLAP database \cite{raasveldt2019duckdb}.

\emph{(c)} In the \emph{analysis stage}, users interact with the collected metrics through a user interface (UI) or an API that connects to the database storing the metrics. Based on the users selecting a target process and two time intervals, baseline and comparison (i.e., application under normal performance and degraded conditions), our approach generates two views. First, a \emph{process dependency graph} of all discovered processes; And second, a \emph{thread dynamics graph} showing interactions between threads and kernel resources at the granularity detailed in Section~\ref{sec:collection}. Using these graphs, users can execute predefined queries to pinpoint state changes and resource contention. This analysis forms the core of the performance degradation analysis process (detailed in Section~\ref{sec:performance-degradation-analysis}) and directly feeds into our \emph{selective thread tracking algorithm} (Algorithm~\ref{algo:selective_tracking}) which systematically traces performance degradation.

}

% Finally, upon collecting the metrics, the analysis stage refers to the insights extracted by these metrics. To facilitate the insight extraction, we provide a user interface through which a user may interact with the metrics collected for a given application, however, by connecting to the database where the metrics are stored, a similar process can be conducted programatically simply by connecting to the same database. In the UI, the user must first specify the database they wish to connect to, followed by specifying the baseline and compare time periods, which respectively represent the time ranges during which the application operated under normal and degraded conditions. The user then selects the process to be analysed which completes the user's input required to perform the automated operations from this stage, i.e., constructing the process dependency graph that the selected process is a part of, and the thread dynamics graph for the selected process, that shows how the threads in the selected application interact with each other and different subsystems, given the granularity of the metrics collected from Section~\ref{sec:collection}. Lastly, the user may execute template (or custom) queries leveraging the input provided by specifying the compare and baseline periods, and the selected application. We leveraged this utility to perform the Selective Thread Tracking process described in Section~\ref{sec:performance-degradation-analysis}.

%-------------------------------------------------------------------------------
\section{Methodology}\label{sec:methodology}
%-------------------------------------------------------------------------------
This section describes our instrumentation methodology, i.\,e., \emph{how} to collect the metrics in Table~\ref{table:metrics}, and how to leverage the metrics for performance degradation analysis.

\subsection{Metric Collection}\label{sec:collection}

\textbf{Scheduling:} A thread's scheduling metrics refers to its time spent in CPU-related states, which are categorized into three main types. \emph{Running} state describes when a thread executes on a CPU core, corresponding to our \emph{runtime} metric in Table \ref{table:metrics}. In the \emph{Idle} state, the thread is not executing; external activity must place it back on a CPU core's runqueue. This state can be further divided into interruptible and uninterruptible states, where the time spent in \verb|TASK_INTERRUPTIBLE| state maps to the \emph{sleep\_time} metric, while the time in the \verb|TASK_UNINTERRUPTIBLE| state corresponds to our \emph{block\_time} metric. A subset of this uninterruptible time is captured by our \emph{iowait\_time} metric, which reflects the time spent in the \verb|TASK_UNINTERRUPTIBLE| state when the \verb|in_iowait| field is set to \verb|1|. Finally, in the \emph{Runnable} state, a thread is ready to run on a core but will not execute while another thread is active. Preemptive schedulers, such as Linux's completely fair scheduler (CFS), manage core time distribution among threads in a runqueue, ensuring fair allocation. The time a thread spends in runqueue without being scheduled is recorded by our \emph{rq\_time} metric.

\textbf{Pipes and Sockets:} In multi-threaded architectures it is common to segregate functionality between the different threads belonging to an application. Sending and receiving data to and from other threads is a fundamental requirement of these architectures to coordinate work. Pipes and sockets are two such mechanisms that facilitate this communication.

When a thread writes to a pipe, data accumulates in a kernel memory buffer. This data is then forwarded in a First-In First-Out (FIFO) fashion to any thread that reads from the same pipe. However, userspace threads use file descriptors (\verb|fd|) to identify the read and write ends of a pipe. File descriptors take the form of integer values that are used to extract the underlying \verb|struct file| pointed at by this descriptor. Multiple file descriptors can point to the same backing resource with varying read/write permissions. For example, there may exist a file descriptor with write permission, and another with read permission, both referring to the same pipe. Therefore, to uniquely identify the backing resource pointed at by the file descriptor, we create a \emph{Backing Resource Identifier} (\emph{BRI}) with a combination of the inode's virtual file system identifiers contained in the \verb|struct file| data structure, namely: \verb|i_ino| representing the \emph{inode identifier}; and \verb|s_dev| denoting superblock device identifier. In the above example with both file descriptors writing to same pipe, the pipe's \verb|i_ino| and \verb|s_dev| will be equivalent, resulting in same \emph{BRI}.

To instrument the time and frequency a thread waits in read/write calls to a FIFO file, given by \emph{pipe\_wait\_\{time,count\}} metrics (see Table~\ref{table:metrics}), we trace the kernel's \verb|vfs_{write,read}| functions. The \verb|struct file| argument in these functions is used to compute the \emph{BRI}, enabling attribution of a thread's wait time to specific pipes. For example, if the \verb|O_NONBLOCK| flag is unset (i.\,e., changing the file descriptor's behavior from non-blocking to blocking mode), our metrics (i.\,e., \emph{pipe\_wait\_\{time,count\}}) account for the time a thread sleeps in its read call while waiting for data to be written on the other end of the pipe.

Sockets, unlike pipes, operate as two-way communication streams with both transmit/receive buffers. For inter-thread communication via sockets, each thread must have a socket mapped to the other, determined by the socket family, type, and protocol. For example, \verb|AF_INET|, \verb|SOCK_STREAM|, and \verb|IPPROTO_TCP| sockets communicate via a TCP connection identified by the protocol 5-tuple:
  $\langle \mathtt{TCP\,protocol},\ \mathtt{source\,IP},\ \mathtt{source\,port},\ \mathtt{destination\,IP},\\ \mathtt{destination\,port} \rangle$.

To monitor the time and frequency a thread waits for socket data reception, denoted by \emph{socket\_wait\_time} and \emph{socket\_wait\_count} metrics (see Table \ref{table:metrics}), we probe kernel's \verb|sock_recvmsg| function. When invoking this function, the kernel passes socket specific data that identifies its respective domain specific source and destination identifiers. We currently support \verb|AF_INET|, \verb|AF_INET6|, and \verb|AF_UNIX| domains. Similarly, on the sending end, we probe domain-specific functions for \verb|AF_INET|, \verb|AF_INET6|, and \verb|AF_UNIX|, i.\,e., \verb|inet_sendmsg|, \verb|inet6_sendmsg|, and \verb|unix*sendmsg| to monitor \emph{socket\_wait\_\{time,count\}} metrics of thread socket data transmission. By collecting these metrics, we can estimate the duration a thread is suspended during interactions with pipe/socket subsystems, highlighting straggling external dependencies that may impede an application's progress.

\textbf{Futex:} Fast userspace mutexes (or futexes) serve as a thread synchronisation mechanism provided by the kernel via the futex system call~\cite{drepper2005futexes}. Two threads synchronize by agreeing on a memory address (\verb|uaddr| parameter) referencing an integer value. For thread sleep requests, the kernel compares the value at \verb|uaddr| with the \verb|val| parameter; if equal, the thread sleeps. To wake sleeping threads, another thread calls futex with the same \verb|uaddr|, specifying \verb|val| threads to wake. Common \verb|val| values are \verb|1| (i.\,e., wake one thread) and \verb|INT_MAX| (i.\,e., wake all threads)~\cite{drepper2005futexes}.

The futex subsystem enables two key use cases. First, it can mediate resource access, limiting the number of threads that can access a particular resource. This functionality underpins various locking mechanisms such as read-write locks, mutexes, and semaphores. Second, futexes can be used as a work scheduling mechanism. In this scenario, a thread is put to sleep when it has no work to process and is awakened by another thread when work becomes available. An example of this is seen in Rust's standard library, which uses futexes in its multi-producer single-consumer (mpsc) message passing implementation. Here, the consumer is parked using futex~\cite{rust-parker} when the message queue is empty and is awakened by a producer when a new message is appended.

To monitor thread interactions with the futex subsystem, we instrument the \verb|sys_enter_futex| (i.\,e., thread entering the futex system call) and \verb|sys_exit_futex| (i.\,e., thread exiting the futex system call) tracepoints. We use the \verb|op| argument to distinguish between futex wake and sleep operations, collecting: total time and frequency a thread sleeps on a specific \verb|uaddr|, represented by \emph{futex\_wait\_time} and \emph{futex\_wait\_count} metrics (see Table~\ref{table:metrics}). The frequency of successful wake operations on a \verb|uaddr|, i.e., waking at least one thread, is given by \emph{futex\_wake\_count} metric. This enables observing an application's lock contention and work scheduling patterns, crucial for understanding how degraded performance propagates across different application threads.

\textbf{Multiplexing IO:} In synchronous programming, with the \verb|O_NONBLOCK| flag unset, a thread executes a system call (e.g., recv) on a single file descriptor (e.g., socket) and must wait for its completion before proceeding to other tasks. High-performance web servers challenged this approach by introducing a scenario where a single thread manages multiple connections concurrently, waiting for incoming data on multiple connections simultaneously \cite{brecht2006evaluating}. For network IO-intensive servers, this approach offers several advantages, including: enhanced CPU cache utilization; decreased memory footprint; and minimized context switching.

The kernel mechanism enabling this new paradigm is multiplexing IO. In essence, a thread specifies which events it wants to consume from a particular file descriptor and adds it to an interest list. It can then wait for all registered \verb|fd|s using a specific system call, which returns when one or more \verb|fd|s have events ready for consumption\footnote{There are other cases that may induce an early return, such as, signal handling and elapsed timers}. 
Linux offers three APIs for this purpose: select, poll, and epoll~\cite{kerrisk2010linux}.

Select and poll share a similar interface where a single system call is used to both express a thread's interests and wait until an event is ready for consumption from file descriptor. For select, we probe entry and exit of \verb|do_select|, while for poll, we probe entry and exit of \verb|do_sys_poll|. Upon returning from these functions, we collect \emph{BRI} of all registered file descriptors and increment each \emph{BRI}'s wait time (e.\,g., \emph{pipe\_wait\_time}, \emph{socket\_wait\_time} metrics in Table \ref{table:metrics}). In essence, a single call to select or poll attributes a thread's wait time to all registered \emph{BRI}s (rather than just one).

In contrast, epoll's API offers three system calls for interacting with the kernel's event poll subsystem. First, \verb|epoll_create| generates an epoll resource and returns a file descriptor (\verb|epfd|) for userspace code to manage its interest list. Second, a thread calls \verb|epoll_ctl| with the \verb|epfd| to add/remove file descriptors from the interest list. Third, a thread invokes \verb|epoll_wait| with the \verb|epfd| to wait for its registered \verb|fd|s.

Given epoll's use of separate system calls to register interest in a \verb|fd|'s events (using \verb|epoll_ctl|) and to wait for registered \verb|fd|s (using \verb|epoll_wait|), our eBPF programs keep a copy of all \verb|fd| \emph{BRI}s registered with each epoll resource. Updating this copy is done, when \verb|fd| events are registered and de-registered using \verb|ep_remove| and \verb|ep_insert| functions.

Furthermore, given that an \verb|epfd| is used by userspace threads to reference an epoll resource, an epoll's \emph{BRI} is derived from the kernel memory address pointing to its \verb|struct eventpoll| data structure. This approach differs from the previous method of calculating a file descriptor's \emph{BRI} because epoll \emph{inodes} are part of the anonymous \emph{inode} filesystem, which reuses the same \emph{inode} for all resources of the same type. Given that multiple threads can wait for the same \verb|epfd|, our first epoll related metric measures the time and frequency a thread spends waiting for a particular epoll \emph{BRI}, represented by \emph{epoll\_wait\_time} and \emph{epoll\_wait\_count} metrics (see Table~\ref{table:metrics}).  Additionally, we account for the time a registered file descriptor is awaited for during an \verb|epoll_wait| system call (i.\,e., \emph{epoll\_file\_wait} metric), attributed to the epoll \emph{BRI} it was registered with.

\textbf{Disk IO:} Given the data persistence requirements of many OLDI applications (e.g., databases and message brokers), their high-level performance metrics can be significantly affected by the maximum throughput achieved when writing data to disk. This is typically because these applications can only confirm message receipt after successful data storage. However, given that disk access is orders of magnitude slower than CPU cache or RAM access, monitoring the time an application's thread spends waiting for disk request completion is crucial. While this information is partially covered by the \emph{iowait\_time} and \emph{block\_time} scheduling metrics, \emph{sector\_count} metric (see Table~\ref{table:metrics}) provides a more comprehensive view of block device activity. It allows for calculating the total in-flight device requests and determining the proportion of these requests belonging to a target application.

% \subsection{Contrasting with Existing Approaches}
\subsection{Application Dynamics Comparison}

\begin{figure}[hbt!]
    \upp\upp\upp
    \centering
    \subfloat[\label{fig:comparison1}]{
      \includegraphics[width=0.3\columnwidth]{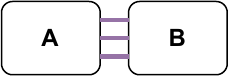} 
    }
    \hspace{0.6cm}
    \subfloat[\label{fig:comparison2}]{
      \includegraphics[width=0.4\columnwidth]{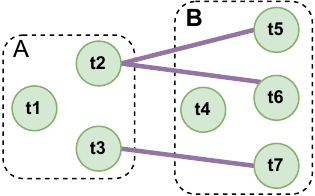} 
    }
    \\ 
    %\upp\upp
    \subfloat[\label{fig:comparison3}]{
    %\upp\upp
      \includegraphics[width=0.7\columnwidth]{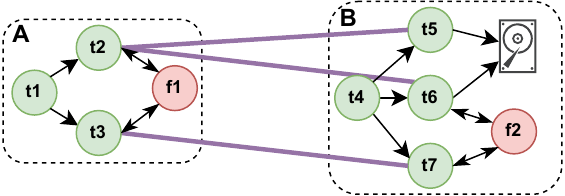} 
    }\\
    \caption{\textbf{Comparison of application interaction observability provided by different approaches, where purple lines denote TCP connections:} (a) Coroot~\cite{coroot}; (b) xCapture~\cite{xCapture}; (c) Our approach. Thread interactions are illustrated by directional arrows.}
   % \upp\upp
\end{figure} 

This section clarifies the distinctions between two existing observability approaches and our metric collection methodology. For comparison, we selected Coroot~\cite{coroot}, a well-established observability tool that collects metrics at the service level (similar to~\cite{pixie}), and xCapture~\cite{xCapture}, which represents methods that collect thread state metrics for TSA. Each approach is qualitatively evaluated using the same setup consisting of two communicating services: an HTTP server (Service \texttt{B}) and a client (Service \texttt{A}). The server exposes two endpoints, one CPU-bound and another disk-bound, while the client periodically issues requests that exercise both.

Figure~\ref{fig:comparison1} presents the information collected by Coroot~\cite{coroot}, which monitors applications at the cgroup level. It captures hardware resource usage per service (e.g., CPU, disk, and memory) and tracks the network connections between them. While this view effectively highlights coarse-grained performance bottlenecks, it provides limited visibility into the internal behavior of individual threads within each process.

Figure~\ref{fig:comparison2} illustrates the data collected by xCapture~\cite{xCapture} for the same application. This approach enhances observability by gathering the metrics required to analyze how each application thread spends its time across the six TSA states. While this provides valuable insight into where execution time is spent within kernel subsystems, it still lacks the context needed to explain why a thread is delayed or where contention originates. To this end, our approach, illustrated in Figure~\ref{fig:comparison3}, attributes each thread’s time to specific kernel resources. By doing so, it captures not only subsystem activity but also the interactions among threads mediated by those resources. This additional context exposes the application’s thread dynamics, i.e, how threads depend on and influence one another through shared locks, pipes, connections, and other synchronization mechanisms. 

As shown in Figure~\ref{fig:comparison3}, threads \texttt{t1} and \texttt{t4} act as work schedulers for their co-located threads, delegating tasks via scheduling futexes (indicated by the directional links between threads). Futexes \texttt{f1} and \texttt{f2} further reveal that threads \texttt{t2} and \texttt{t3}, and \texttt{t6} and \texttt{t7}, are contending for shared resources. Furthermore, threads \texttt{t5} and \texttt{t6}, as seen in the thread dynamics graph, show these threads are responsible for performing the disk-intensive tasks. Consequently, our approach exposes not only where time is spent, but also how performance degradation propagates across threads and services through these kernel-level interactions. This visibility allows analysts to move beyond coarse-grained metrics and pinpoint the specific thread–resource relationships responsible for degraded performance, making our approach particularly effective for analyzing multi-threaded applications.

% {\color{blue}However, while our approach provides finer-system granularity, but does not replace traditional APM tools with flame graphs, distributed tracing, and request-level performance metrics. Instead, it complements them adding system-level insight into how software interacts with kernel resources during execution.}

% {\color{blue}
% However, while our approach provides added system granularity, information provided by common Application Performance Monitoring (APM) tools (e.g. flame graphs, distributed tracing, and request-level performance metrics), continues to play a critical role at different stages of performance degradation analysis. Accordingly, rather than positioning our approach as a replacement for these tools, our goal is to illustrate how it complements existing APM solutions by providing an additional layer of insight into how software components interact with the underlying system during execution.
% }

\subsection{Performance Degradation Analysis with Thread Dynamics}
\label{sec:performance-degradation-analysis}
% \begin{figure}[h]
% \centering
% \includegraphics[width=0.45\columnwidth]{figs/thread-dynamics-1.pdf}
%     \caption{Illustration of an application's thread dynamics derived from resource-level metric collection.} 
% %\upp\upp\upp
% \label{fig:thread-dynamics}
% \end{figure}

Thread State Analysis (TSA) provides a strong foundation for diagnosing performance degradation. It relies on observing shifts in thread state metrics and identifying the subsystems constraining application performance. However, by collecting thread state metrics at resource-level granularity, our approach goes further: it reveals not only which subsystem is causing the slowdown but also how the degradation propagates across interacting threads. This section defines a systematic procedure for analyzing performance degradation in online data-intensive applications using this extended form of TSA.

\begin{wrapfigure}{R}{0pt}
  \centering
 % \upp\upp
  \includegraphics[width=.5\columnwidth]{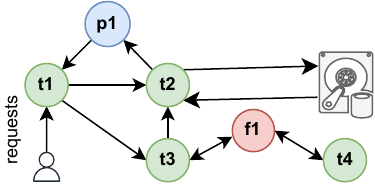}
	 \caption{Illustration of an application's thread dynamics derived from resource-level metric collection.}
  \label{fig:thread-dynamics}
  %\upp\upp
\end{wrapfigure}
\emph{\bf Example:} Figure~\ref{fig:thread-dynamics} illustrates an example of an application thread dynamics graph constructed from the collected metrics. Each arrow between two threads, such as \texttt{t1}$\rightarrow$\texttt{t2}, represents a userspace schedule relationship where thread \texttt{t1} schedules \texttt{t2} via a futex resource. The futex node \texttt{f1} corresponds to a synchronization primitive used to coordinate access to shared resources. Since both threads wait and wake on the same futex, the edges connecting them to \texttt{f1} are bi-directional. Similarly, pipe \texttt{p1} connects a writing thread (\texttt{t2}) to a reading thread (\texttt{t1}), forming a dependency that indicates data transfer through an IPC channel. As such, the graph exposes the propagation paths through which a thread’s degraded performance may influence other threads within the application.

\begin{algorithm}
\scriptsize
\SetAlgoLined
\KwIn{target application, target metric time series}
\KwOut{Set of candidate threads $T_{track}$ and their anomalous metrics}
\textbf{Initialize:} $T_{track} \gets \emptyset$\;
\BlankLine
$T_{entry} \gets$ threads with \texttt{socket\_wait\_\{time,count\}} activity for target application\;
$T_{track} \gets T_{track} \cup T_{entry}$\;
$T_{seen} \gets \emptyset$\;
\Repeat{no new threads are added to $T_{track}$}{
    $M_{flagged} \gets \emptyset$\;
    \ForEach{thread $t \in T_{track} \setminus T_{seen}$}{
        \ForEach{metric $m \in \text{TSA states}$ for thread $t$}{
            \If{distribution of $m$ shifts during KPI degradation}{
                $M_{flagged} \gets M_{flagged} \cup \{(t, m)\}$\;
            }
        }
        $T_{seen} \gets T_{seen} \cup \{t\}$\;
    }
    $T_{new} \gets \emptyset$\;
    \ForEach{$(t, m) \in M_{flagged}$}{
        \If{$m$ is a \texttt{futex}, \texttt{pipe}, or \texttt{socket} metric}{
            $BRI \gets$ Backing Resource Identifier from metric $m$\;
            $T_{counterparts} \gets$ all threads that interacted with $BRI$ (e.g., via \texttt{futex\_wake}, pipe write, etc.)\;
            $T_{new} \gets T_{new} \cup T_{counterparts}$\;
        }
    }
    $T_{track} \gets T_{track} \cup T_{new}$\;
}
\Return $(T_{track}, M_{flagged})$\;
\caption{Selective Thread Tracking}
\label{algo:selective_tracking}
\end{algorithm}

% Building on this graph, our performance degradation analysis employs \emph{selective thread tracking} procedure. It begins by identifying entrypoint threads, those engaged in external communication, using the \emph{socket\_wait\_{time,count}} metrics for IPv4/IPv6 sockets. Each of these threads is then analyzed based on its thread state metrics to detect distribution shifts in any subsystem that align with the observed performance degradation. When such a shift is identified, we further examine the affected subsystem’s resources to pinpoint those likely responsible for the degradation. The application’s thread dynamics graph (e.g., Figure~\ref{fig:thread-dynamics}) is then leveraged to identify other threads interacting with the implicated resource, which are subsequently analyzed using the same procedure. This iterative process continues until no additional threads exhibit behavior correlated with the application’s degraded performance. Ultimately, this process identifies a subset of application threads and resources at the center of the degraded performance, along with how they directly and indirectly interact to constrain overall performance.

\emph{\bf Selective Thread Tracking:} Building on the thread dynamics, our performance degradation analysis employs a \emph{selective thread tracking} procedure outlined in Algorithm~\ref{algo:selective_tracking}. The algorithm takes a target application and time-series of its performance metrics (e.g., response time or throughput) as input, and outputs candidate threads with anomalous metrics correlated to performance degradation. 
{\color{black}
The algorithm begins by identifying entry‑point threads ($T_{entry}$), defined as threads that  form  the application’s interface to incoming service requests, by detecting any thread that reads from or writes to IPv4/IPv6 sockets using the socket metrics defined in Section~\ref{sec:collection}.
% The algorithm begins by identifying entry-point threads ($T_{entry}$), defined as the threads that form the application’s interface to incoming service requests. This requires identifying any thread that reads from or writes to IPv4 or IPv6 sockets via the socket metrics from Section~\ref{sec:collection}.
}
These threads are added to the initial tracking list ($T_{track}$). Next, we perform correlation analysis which analyzes the time-series data for all metrics in Table~\ref{table:metrics} for all threads that have not yet been analyzed in $T_{track}$. Metrics with significant statistical distribution shifts during performance degradation periods suggested by methods such as {\color{black} Mann Whitney U, Kolmogorov-Smirnov,} or change point detection, are flagged for further analysis.

For each flagged IPC metric (\emph{futex, pipe, socket*}), the algorithm utilizes their granular resource identification. When thread $t$ shows anomalous wait time on a specific resource (identified by its \emph{BRI}, defined in Section~\ref{sec:collection}), all threads interacting with the same \emph{BRI} are identified as counterpart threads ($T_{new}$) and added to $T_{track}$ if not already in the list. These steps iterate until no new unique threads are added to $T_{track}$, allowing us to capture all inter-dependent threads contributing to performance degradation.

The algorithm concludes by returning the final set of inter-dependent threads $T_{track}$ and their flagged metrics ($M_{flag}$) associated with specific kernel resources. This methodology effectively narrows the diagnostic search space by tracing dependency chains between threads, focusing exclusively on those directly or indirectly interacting with outward-facing threads. 
{\color{black}
However, the algorithm optimistically assumes degradation propagates toward an entry‑point thread. If no resource contention is found after exhausting the metric search space in Algorithm~\ref{algo:selective_tracking}, a computationally expensive full search across all process metrics can be performed.
% It is important to note, however, that this algorithm optimistically assumes the existence of a degradation path that propagates toward an entry-point thread. If no significant signs of resource contention are found and the metric search space defined by Algorithm~\ref{algo:selective_tracking} is exhausted, we plan to extend this approach with a more comprehensive, albeit computationally more expensive, algorithm that performs a full search over all metrics associated with the process.
}
The selective thread tracking algorithm has an over all time 
%and space 
complexity of  $O(k \cdot n )$ , where \emph{n, k} denote system threads and \emph{BRI} dependencies for each thread, respectively. Given that we set an upper bound on the total amount of BRIs tracked per thread, the algorithm scales linearly with the number of threads involved in performance degradation.

%-------------------------------------------------------------------------------
\section{Experiments}
\label{sec:experiments}
%-------------------------------------------------------------------------------
This section presents experimental setup, selected applications, induced workload and performance degradation scenarios to evaluate the diagnostic capability of our method.

%\paragraph{\textbf{Experimental Setup}} 
%\subsection{Experimental Setup}
\textbf{Experimental Setup:} We implemented our method as a Rust executable that leverages \emph{libbpf} to interface with \emph{bpf}. It cyclically gathers kernel statistics \emph{every second} using our custom eBPF programs, transforms them into metrics listed in Table \ref{table:metrics}, and persists them for further analysis. The metric collector's complete source code and this Section's reproducibility artifact\footnote{Artifact is also included in the two-page Artifact Description (AD)/Artifact Evaluation (AE) appendix, at the end of the paper.} is available in an \emph{online} \verb|GitHub| repository~\cite{repo}. We executed our experiments on \verb|i7-11850H Intel| server @ \SI{2.50}{\giga\hertz} running  Ubuntu 22.04 LTS, Jammy Jellyfish (\verb|x86_64|) OS with \verb|6.8.12| Linux kernel, $8$ CPU cores, $2$$\times$\SI{16}{\giga\byte} \verb|DDR4| RAM, and \SI{1}{\tera\byte} SSD disk.

%\paragraph{\textbf{Selected Applications}}
%\subsection{Selected Applications}
\textbf{Selected Applications:} We selected \emph{six} online data-intensive applications for validation. Our application set includes a relational and a NoSQL database, \emph{MySQl} and \emph{Apache Cassandra}. We also selected \emph{Redis} which fits into the category of in-memory Key-value datastores, while \emph{Kafka} is a message broker middleware. To incorporate a broader spectrum of applications, our evaluation set also includes a microservice-based benchmark application \emph{TeaStore}~\cite{von2018teastore}. Finally, we also evaluate a machine learning inference (ML-inference) service wrapped around a sentiment analysis model~\cite{perez2021pysentimiento} in a \emph{FastAPI web-app}, deployed using a \verb|Python-based| webserver gateway interface \emph{gunicorn}.

Most of these selected applications span across a wide range of domains and have been widely used in previous literature\cite{grohmann2019monitorless, xing2023hybrid, palit2016demystifying, wang2023characterizing}. Additionally, the chosen applications provide the variability with regards to combining and utilizing different threading models, inter-process communication (IPC) methods, and user-space virtual machines. {\color{black}
Regarding threading models, \emph{MySQl} uses a thread pool, \emph{ML-inference}  a process pool, and \emph{Kafka}, \emph{Teastore}, \emph{Cassandra}, and \emph{Redis}  use asynchronous multiplexing I/O for handling client requests.
% For instance, with regards to threading models to handle client requests: \emph{MySQl} leverages a thread pool; \emph{ML-inference} a process pool; and \emph{Kafka}, \emph{Teastore}, \emph{Cassandra}, and \emph{Redis} asynchronous multiplexing IO
} 
Concerning userspace virtual machines, \emph{Kafka} and \emph{Cassandra} operate on the Java Virtual Machine (JVM), while the ML-inference service runs on the Python Virtual Machine. Lastly, with regards to IPC, \emph{Kafka} and \emph{Teastore} leverage pipes to communicate between threads, while futex mechanisms are used by all applications for synchronisation and work scheduling activities.

%\paragraph{\textbf{Workload Generation}} 
%\subsection{Workload Generation}
\textbf{Workload Generation:} We used YCSB~\cite{cooper2010benchmarking} and TPCC~\cite{leutenegger1993modeling} benchmarks to generate read- and update-intensive workloads for \emph{MySQL}, while for \emph{Cassandra}, only YCSB was used. For \emph{Redis}, we used the official ~\cite{redis-benchmark} and the memtier~\cite{redis-memtier} benchmarks to saturate the server. Workload for the ML-inference server and \emph{Teastore} applications  was generated using the Locust~\cite{locust} framework. We applied the same workload pattern for both cases, varying load between \SIrange[range-phrase=\,--\,]{2}{60}{} users. For the ML-inference server specifically, the server is queried about the sentiment of tweets we extract from a twitter dataset \cite{twitter-dataset}, whereas \emph{Teastore}'s requests are generated by the request generator provided in their official repository \cite{von2018teastore}. Finally, for \emph{Kafka} we generated a variable workload pattern by creating a custom producer with an increasing production rate of up to \SI{450,000}{} events per second.

%\paragraph{\textbf{Degradation Scenarios}} 
%\subsection{Degradation Scenarios}
\textbf{Degradation Scenarios:} We utilized a combination of two approaches to degrade the selected applications performance. The first approach used benchmarks to increase the application's load to the point of saturation, impairing its performance, similar to ~\cite{gregg2013thinking}. The second approach induced a combination of CPU, disk and lock contention to interfere with the application's activity and hinder its performance. CPU contention involved either underprovisioning CPU resources (e.g. \emph{Teastore} and \emph{Cassandra}) or competing for same core on-cpu time (e.g. \emph{Redis}). As for disk contention, inspired by stress-ng's \cite{stressng} \verb|hdd| stress procedure, we start 80 threads that execute synchronous writes to disk (\verb|O_SYNC| flag set) at their maximum achievable rate, to compete for the underlying disk resource. Lastly, with regards to lock contention, we induced this type of contention on MySQL due to its ACID compliance \cite{mysql-acid}. We performed this by executing two distinct YCSB workloads reading from and writing to the same dataset.

%\paragraph{\textbf{Target Performance Metrics}} 
%\subsection{Target Performance Metrics}
\textbf{Target Performance Metrics:} To evaluate the instrumented metrics' robustness to different high-level KPIs, we vary the target performance metrics in each experiment. In case of \emph{MySQL}, we use two target metrics: $95^{th}$ percentile \emph{response time} (seconds) and \emph{throughput} (requests per second), which measures \emph{MySQL's} performance as seen by the YCSB and TPCC workloads. \emph{Cassandra's} performance is measured using median \emph{response time}, while \emph{Kafka} uses the total production rate (i.e. \emph{throughput}) in events per second, as a proxy of its performance. Similarly, we used $95^{th}$ percentile \emph{response time} as a target metric to estimate ML-inference server, \emph{Redis} and \emph{Teastore} application performance.

%-------------------------------------------------------------------------------
\section{Results}\label{sec:results}
%-------------------------------------------------------------------------------
\begin{figure}[t]
    \centering
   % \upp\upp\upp\upp
    \subfloat[\label{fig:mysql:throughput}]{
      \includegraphics[width=0.3\columnwidth]{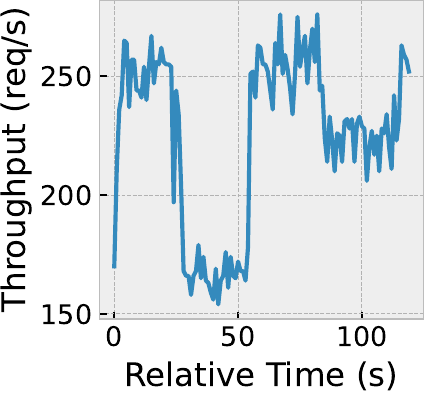}
    }
    \subfloat[\label{fig:mysql:latency}]{
      \includegraphics[width=0.3\columnwidth]{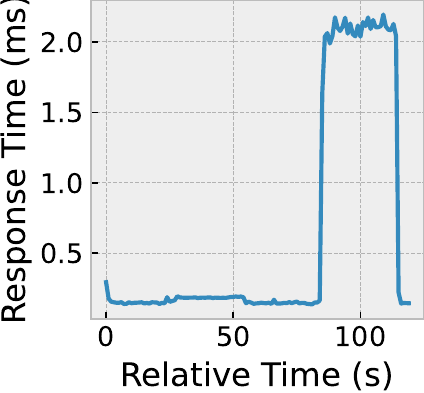}
    }
    \subfloat[\label{fig:mysql:device_reqs}]{
      \includegraphics[width=0.32\columnwidth]{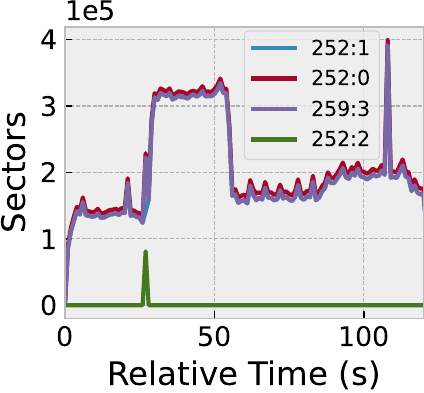}
    }\\  
  %  \upp\upp 
    \subfloat[\label{fig:mysql:device_share}]{
    \upp\upp
      \includegraphics[width=0.3\columnwidth]{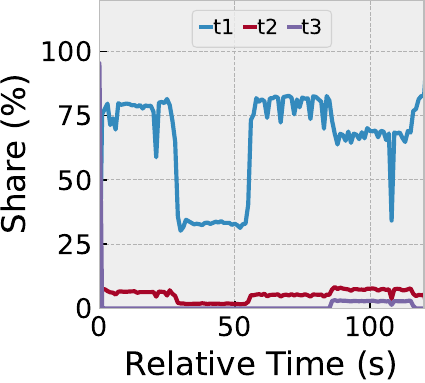}
    }
    \subfloat[\label{fig:mysql:lock_futex_wait}]{
 %   \upp\upp
      \includegraphics[width=0.3\columnwidth]{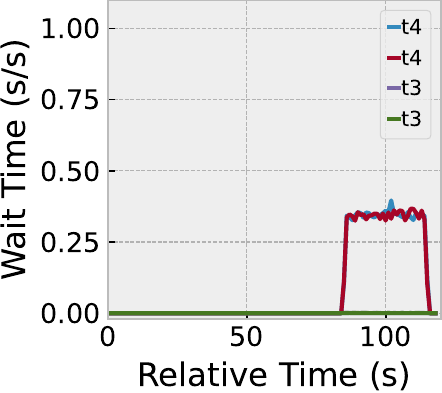}
    }
    \subfloat[\label{fig:mysql:lock_futex_wake}]{
  %  \upp\upp
      \includegraphics[width=0.29\columnwidth]{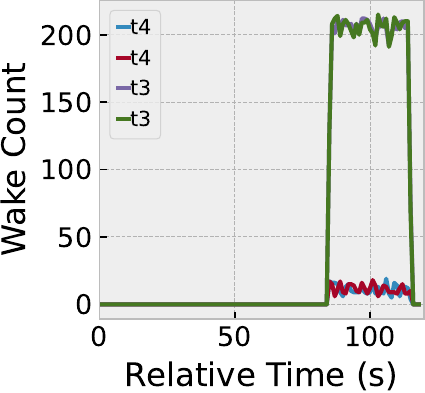}
    }\\ 
%    \upp\upp 
    \subfloat[\label{fig:mysql:disk_sched_iowait}]{
%    \upp\upp
      \includegraphics[width=0.3\columnwidth]{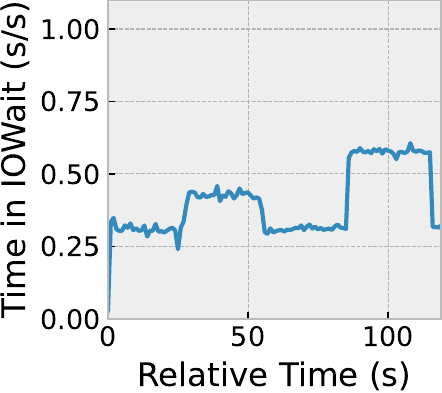}
    }
    \subfloat[\label{fig:mysql:disk_futex_wait}]{
%    \upp\upp
      \includegraphics[width=0.3\columnwidth]{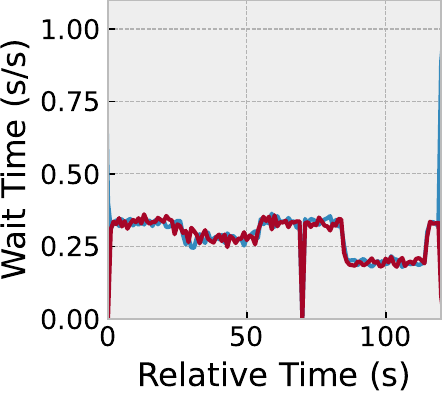}
    }
    \subfloat[\label{fig:mysql:disk_futex_wake}]{
%    \upp\upp
      \includegraphics[width=0.29\columnwidth]{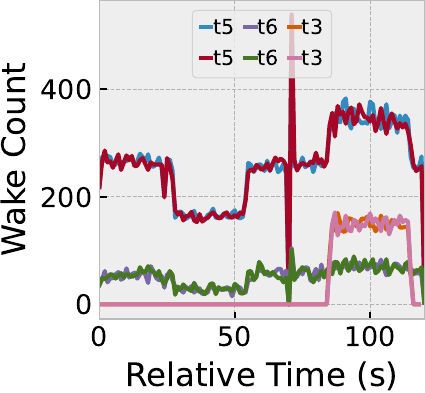}
    }
    % \\
    % \subfloat[\label{fig:mysql:thread_dynamics}]{
    %  %\upp\upp
    %   \includegraphics[width=0.8\linewidth]{figs/mysql.thread_dynamics.pdf}
    % }
    %\vspace*{-0.1cm}
   % \upp
    \caption{\textbf{MySQL target metric observation and performance degradation deconstruction:} 
    (a) TPCC workload throughput; (b) YCSB read-intensive workload $95^{th}$ percentile response time; (c) Sector requests per device (identified by their \texttt{major:minor|values}; (d) Device request share for MySQL threads; 
    (e) Per thread futexes' (\texttt{f1, f2}) wait time; (f) Per thread futexes' (\texttt{f1, f2}) wake activity; (g) Thread \texttt{t2}'s \emph{iowait\_time}; (h) Thread \texttt{t2}'s futexes' (\texttt{f3, f4}) wait time; (i) Per thread futexes' (\texttt{f3, f4}) wake activity. {\color{blue}
    %(j) MySQL thread dynamics for the degradation path, where blue threads denote entry-point threads.
    }
    } 
    \upp
\end{figure}

This section demonstrates how the \emph{eBPF-based} metrics defined in Table \ref{table:metrics} and collected using methodology in Section~\ref{sec:collection} are used to analyze application performance degradation, under the workload patterns and scenarios from Section~\ref{sec:experiments}, through the procedure outlined in Section \ref{sec:performance-degradation-analysis}. Our goal is to infer not only the root cause of degradation but also how it propagates across threads as a result of the application’s thread dynamics. By correlating thread state shifts with interactions over shared kernel resources, we can trace how localized contention or blocking cascades through dependent threads, ultimately constraining user-facing performance.

Degradation analysis results for each application introduced in this section are grouped by their primary resource constraint. For clarity, threads, futexes, epoll instances, and pipes are denoted as \texttt{t}, \texttt{f}, \texttt{e}, and \texttt{p}, respectively, instead of using raw kernel identifiers. Each case highlights only the most relevant metrics and interactions necessary to explain the observed performance degradation. 
% {\color{blue}
% When thread dynamics critically contribute to degradation, we also include the application’s thread dynamics graph.
% % Additionally, when thread dynamics play a critical role in the observed degradation, we include the thread dynamics graph for the application under analysis.
% }
All experimental data and analysis artifacts are available in an online public repository~\cite{repo}.

\subsection{Disk Constrained}
\subsubsection{MySQL} \label{sec:mysql} \textbf{Experiment Scenario \& Target Metric Observation:} 
We executed TPCC and YCSB read-intensive workloads over \SI{120}{\second}. Two controlled interventions were introduced during this period. The first intervention, beginning \SI{25}{\second} into execution and lasting \SI{30}{\second}, injected a disk contention workload. The second intervention, starting at \SI{85}{\second} and lasting \SI{30}{\second}, ran a YCSB update-intensive workload that induced both lock contention with the YCSB read workload and disk contention with the TPCC benchmark due to update operations.

Figures~\ref{fig:mysql:throughput} and~\ref{fig:mysql:latency} present TPCC throughput and YCSB $95^{th}$ percentile latency across the experiment. During the first intervention, TPCC throughput declines sharply, while YCSB latency remains unaffected, consistent with TPCC’s disk-intensive behavior versus YCSB’s in-memory reads. The second intervention, in contrast, degrades both metrics, confirming interference between the concurrent read and update workloads.

% \begin{wrapfigure}{R}{0pt}
%   \centering
%   \upp\upp\upp
%   \includegraphics[width=.55\columnwidth]{figs/thread-dynamics-mysql-revised.pdf}
% 	 \caption{{\color{black}MySQL thread dynamics for the degradation path: blue threads denote entry-point threads.}}
%   \label{fig:mysql:thread_dynamics-rev}
%   \upp
% \end{wrapfigure}

\begin{figure}[hbt!]
  \centering
 %\upp\upp
  \includegraphics[width=.7\columnwidth]{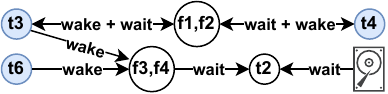}
	 \caption{{\color{black}MySQL thread dynamics for the degradation path: blue threads denote entry-point threads.}}
  \label{fig:mysql:thread_dynamics-rev}
 % \upp
\end{figure}

\textbf{Deconstructing Performance Degradation:} 
Following the procedure in Section~\ref{sec:performance-degradation-analysis}, we first identify the entry-point threads \texttt{t3}, \texttt{t4}, and \texttt{t6} using their \emph{socket\_wait\_{time,count}} metrics.
{\color{black}
Threads \texttt{t3} and \texttt{t6} then wake \texttt{t2} via futexes \texttt{f3} and \texttt{f4} (Fig.~\ref{fig:mysql:disk_futex_wake}). The decrease in \texttt{t2}'s \emph{futex\_wait\_time} (Fig.~\ref{fig:mysql:disk_futex_wait}), alongside a rise in its \emph{iowait\_time}, signals disk I/O contention in both interventions. Figure~\ref{fig:mysql:device_reqs} shows that during the first intervention MySQL competes with co‑located processes for disk access, whereas the second intervention reveals increased futex activity (Fig.~\ref{fig:mysql:disk_futex_wake}) indicating an additional client, handled by \texttt{t3} is contending for the same disk resource.

The second degradation path involves lock contention between YCSB read and update-intensive  workloads. Thread \texttt{t4}'s \emph{futex\_wait\_time} rises sharply (Fig.~\ref{fig:mysql:lock_futex_wait}), and mutual wake activity between \texttt{t3} and \texttt{t4} (Fig.~\ref{fig:mysql:lock_futex_wake}) confirms synchronization over shared futexes. Since \texttt{t3} handles updates and \texttt{t4} handles reads, this lock contention directly contributes to YCSB’s latency degradation. 

Overall, the thread dynamics (Fig.~\ref{fig:mysql:thread_dynamics-rev}) illustrate how localized lock and I/O contention propagates through futex‑mediated scheduling and shared I/O channels, ultimately degrading application‑level throughput and latency.
}

\begin{figure}[t]
    \centering
  %  \upp\upp\upp\upp
    \subfloat[\label{fig:kafka:throughput}]{
      \includegraphics[width=0.3\columnwidth]{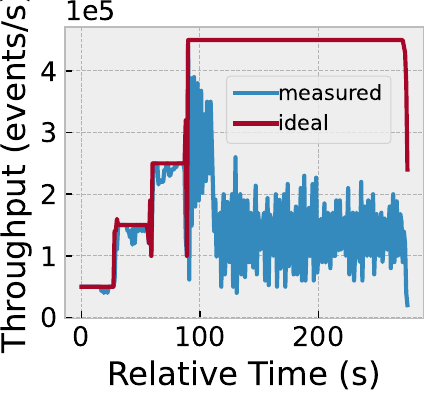}
    }
    \subfloat[\label{fig:kafka:runtime}]{
      \includegraphics[width=0.3\columnwidth]{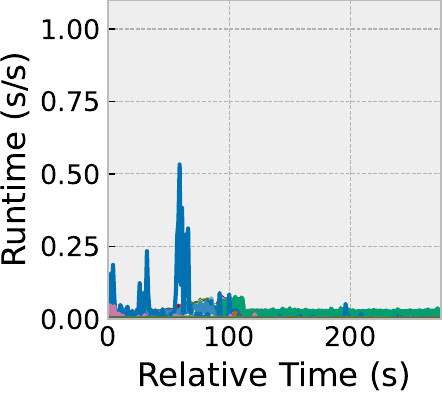}
    }
    \subfloat[\label{fig:kafka:block_time}]{
      \includegraphics[width=0.28\columnwidth]{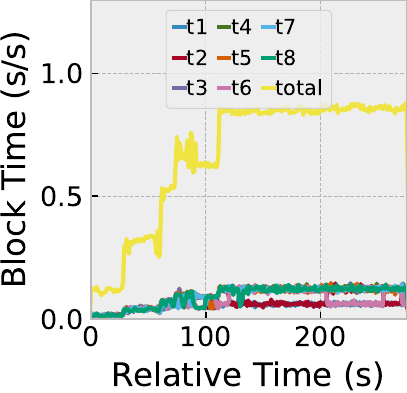}
    }\\ 
    %\upp\upp
    % \subfloat[\label{fig:kafka:futex_wait}]{
    %  %\upp\upp
    %   \includegraphics[width=0.25\linewidth]{graphs/kafka/futex_wait_time.pdf}
    % }
    % \subfloat[\label{fig:kafka:thread_sched}]{
    %  %\upp\upp
    %   \includegraphics[width=0.24\linewidth]{graphs/kafka/sched_activity_633923.pdf}
    % }
    % \subfloat[\label{fig:kafka:futex_wake}]{
    %  %\upp\upp
    %   \includegraphics[width=0.24\linewidth]{graphs/kafka/futex_wake_count.pdf}
    % }\\ 
   % \upp\upp
    \subfloat[\label{fig:kafka:external_wait}]{
   %  \upp\upp
      \includegraphics[width=0.3\columnwidth]{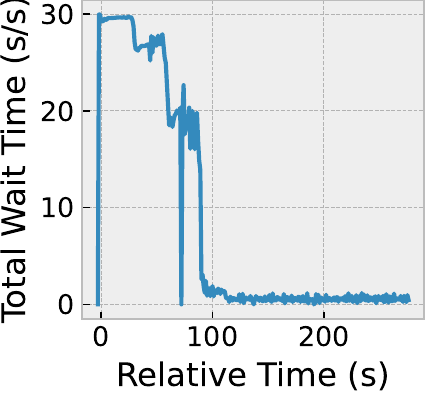}
    }
    \subfloat[\label{fig:kafka:epoll_wait}]{
 %    \upp\upp
      \includegraphics[width=0.3\columnwidth]{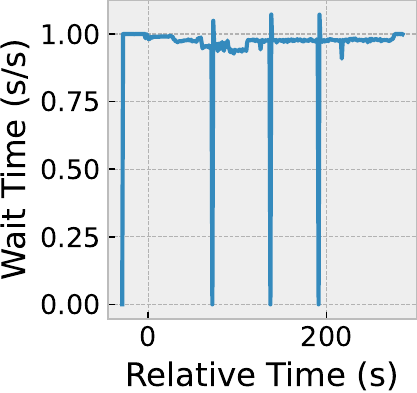}
    }
    \subfloat[\label{fig:kafka:pipe_write}]{
 %    \upp\upp
      \includegraphics[width=0.3\columnwidth]{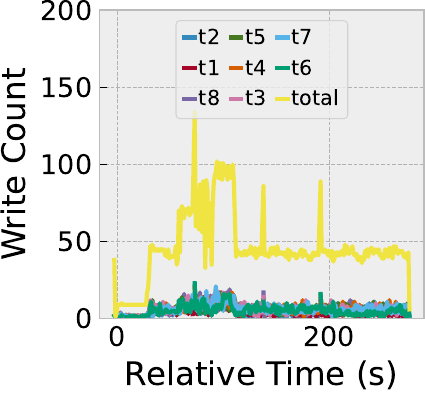}
    }
%    \upp
    % \\
    % \subfloat[\label{fig:kafka:thread_dynamics}]{
    %  %\upp\upp
    %   \includegraphics[width=0.8\linewidth]{figs/kafka.thread_dynamics.pdf}
    % }
    %  %\vspace*{-0.2cm}
    \cprotect\caption{\textbf{Kafka target metric and performance degradation deconstruction:} (a) Ideal/measured throughput; (b) Kafka threads' \emph{runtime}; (c) Kafka threads' \emph{block\_time}; {\color{black}(d) Combined connection wait time for epoll \verb|e1|; (e) Thread \verb|t9| epoll \verb|e1| wait time; (f) Pipe \verb|p1| write activity. 
    %(g) Kafka thread dynamics for the degradation path, where blue threads denote entry-point threads.
    }} 
    \upp\upp
\end{figure}

\subsubsection{Kafka} \label{sec:kafka}
\begin{wrapfigure}{R}{0pt}
  \centering
  %\upp\upp\upp\upp
  \includegraphics[width=.5\columnwidth]{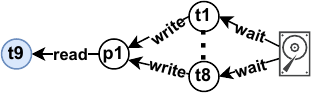}
	 \caption{{\color{black}Kafka thread dynamics for the degradation path: blue threads denote entry-point threads.}}
  \label{fig:kafka:thread_dynamics-rev}
 % \upp
\end{wrapfigure}
\textbf{Experiment Scenario \& Target Metric Observation:}
For the \emph{Kafka} experiment, we progressively increased the production load until the broker could no longer sustain the incoming rate. Figure~\ref{fig:kafka:throughput} compares the ideal and actual production rates, showing that the system begins to falter around an ideal rate of \SI{450,000}{} events per second, where the observed production rate drops by approximately $55$–$77$\% below target.

\textbf{Deconstructing Performance Degradation:}
We start by identifying the entrypoint threads responsible for external communication (\texttt{t9-t11}) using the \emph{socket\_wait\_{time,count}} and \emph{epoll\_wait\_{time,count}} metrics. These threads each manage dedicated epoll resources to handle incoming producer requests and send acknowledgments. During the degradation phase, Figure~\ref{fig:kafka:epoll_wait} shows that thread \texttt{t9} spends approximately 100\% of its time waiting on its epoll backing resource identifier (BRI). Combined with Figure~\ref{fig:kafka:external_wait}, this suggests that \texttt{t9} is awaiting for another type of resource than client connections through epoll \texttt{e1}. Examining the resources registered with \texttt{e1} reveals a pipe \texttt{p1} that shows a sharp decrease in write activity (Figure~\ref{fig:kafka:pipe_write}) throughout the same time period as kafka saturated.
The threads responsible for writing to this pipe (\texttt{t1-t8}) are part of a Kafka worker pool, whose combined time in uninterruptible wait, saturated at the same moment as Kafka's throughput degraded.

{\color{black}
Mapping these interactions to the thread dynamics graph (Fig.~\ref{fig:kafka:thread_dynamics-rev}) reveals a clear propagation path: worker threads (\texttt{t1-t8}) saturate in \emph{block\_time}, delaying writes to the internal pipe; this stalls the pipe’s consumer, thread \texttt{t9}, which then delays replies to client requests.
}

\begin{figure}[t]
    \centering
 %   \upp\upp\upp\upp
    \subfloat[\label{fig:cassandra:response_time}]{
      \includegraphics[width=0.3\columnwidth]{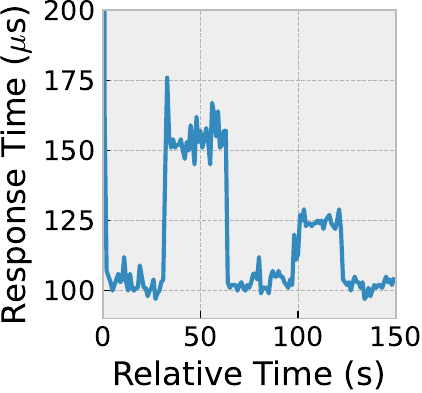}
    }
    \subfloat[\label{fig:cassandra:first_rqtime}]{
      \includegraphics[width=0.31\columnwidth]{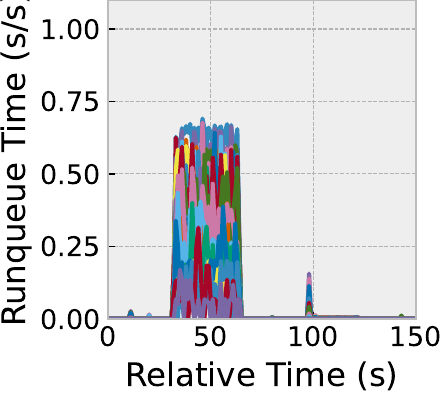}
    }
    %\upp\upp
    \subfloat[\label{fig:cassandra:second_runtime}]{
    %\upp\upp
      \includegraphics[width=0.31\columnwidth]{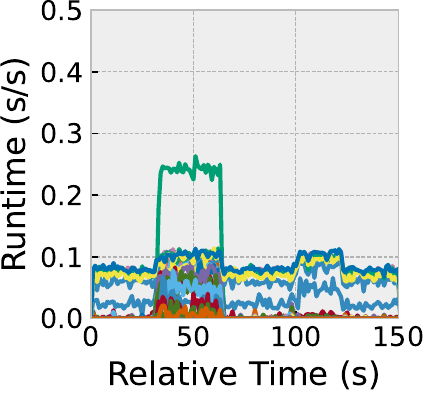}
    }\\
  %  \upp\upp
    \subfloat[\label{fig:cassandra:iowait_time}]{
  %  \upp\upp
      \includegraphics[width=0.31\columnwidth]{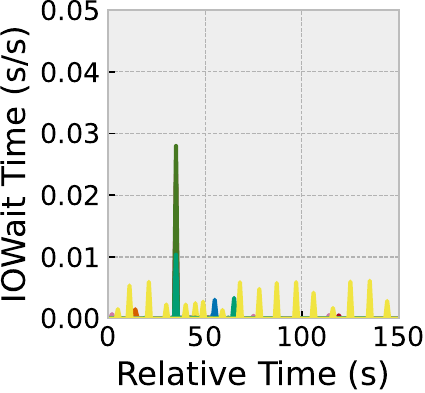}
    }
    %\upp\upp\upp
    \subfloat[\label{fig:cassandra:device_reqs}]{
  %  \upp\upp
      \includegraphics[width=0.3\columnwidth]{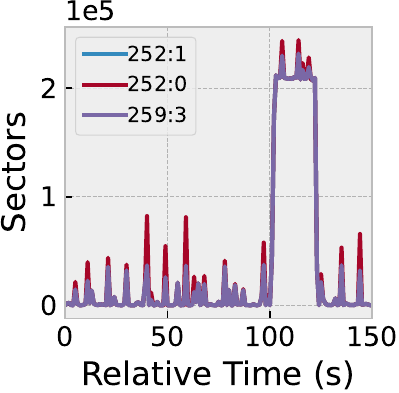}
    }
    \subfloat[\label{fig:cassandra:device_share}]{
 %   \upp\upp
      \includegraphics[width=0.31\columnwidth]{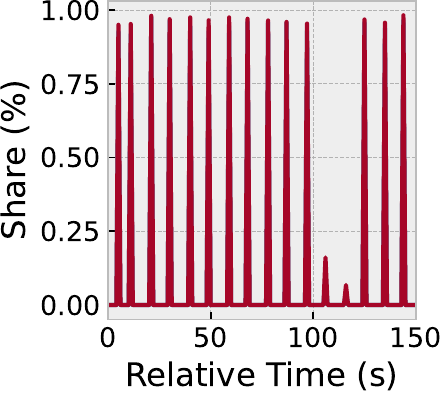}
    }
 %   \upp
    \caption{\textbf{Cassandra target metric observation and performance degradation deconstruction:} (a) YCSB update-intensive workload median response time; (b) Cassandra threads' \emph{rq\_time} throughout the first intervention; (c) Threads' \emph{runtime} throughout the second intervention; (d) Thread \emph{iowait\_time} throughout the second intervention; (e) Sector requests per device; (f) Device request share for Cassandra threads.} 
  %  \upp\upp\upp
\end{figure}

%\paragraph{\textbf{Cassandra}}
\subsubsection{Cassandra}\textbf{Experiment Scenario \& Target Metric Observation:} For the \emph{Cassandra} experiment, we execute a YCSB update-intensive workload while measuring the median response time (Figure~\ref{fig:cassandra:response_time}) as the target metric. Two interventions are introduced. The first, at \SI{33}{\second}, launches a \SI{30}{\second} YCSB read-intensive workload targeting the same dataset to induce lock contention. The second, from \SI{100}{\second} to \SI{125}{\second}, introduces a disk contention workload to interfere with Cassandra’s disk operations by saturating the underlying device.

\textbf{Deconstructing Performance Degradation:} For first scenario, Figure~\ref{fig:cassandra:first_rqtime} shows a significant increase in \emph{rq\_time} of the request-facing threads which led to Cassandra's degraded performance. Contrary to our expected cause of degradation, CPU contention, and not locking, was the cause of degradation.

In the second intervention, Figure \ref{fig:cassandra:response_time} shows a slight increase in \emph{Cassandra's} median response time. Unexpectedly, Figure \ref{fig:cassandra:second_runtime} shows increased runtime despite the applied load remaining constant, which suggests the intervention led to memory contention, that manifested as an increase in runtime due to additional stalled CPU cycles. To further investigate why disk isn't the cause of performance degradation, Cassandra's IO path shown in Figures~\ref{fig:cassandra:iowait_time}, \ref{fig:cassandra:device_reqs}, and \ref{fig:cassandra:device_share} show that data is persisted to disk asynchronously, and, consequently, does not directly interfere with the client requests. Further investigating Cassandra's documentation~\cite{cassandradocs} reveals that with \texttt{commitlog\_sync}  set to \verb|periodic| and \verb|commit_log_sync_period| to \verb|10000|, Cassandra waits \SI{10000}{\milli\second} before \emph{fsync}'ing the commit log, explaining the observed behavior.

\subsection{CPU Constrained}

\begin{figure}[t]
    \centering
  %  \upp\upp\upp
    \subfloat[\label{fig:ml:load}]{
      \includegraphics[width=0.3\columnwidth]{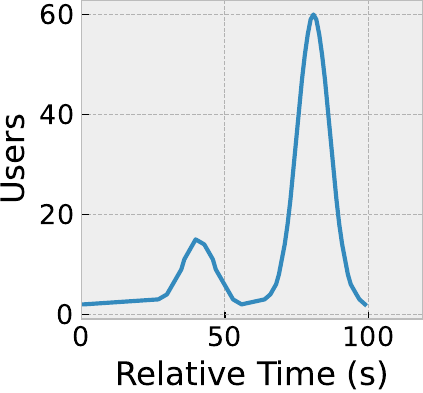}
    }
    \subfloat[\label{fig:ml:target}]{
      \includegraphics[width=0.3\columnwidth]{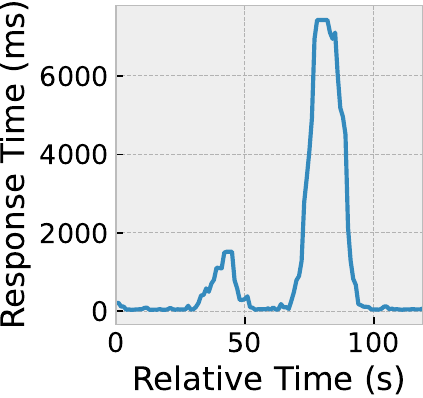}
    }
    %  \subfloat[\label{fig:ml:runtime}]{
    %   \includegraphics[width=0.25\linewidth]{graphs/ml-inference/runtime.pdf}
    % }\\ 
 %   \upp\upp
    \subfloat[\label{fig:ml:rqtime}]{
  %  \upp\upp
      \includegraphics[width=0.31\columnwidth]{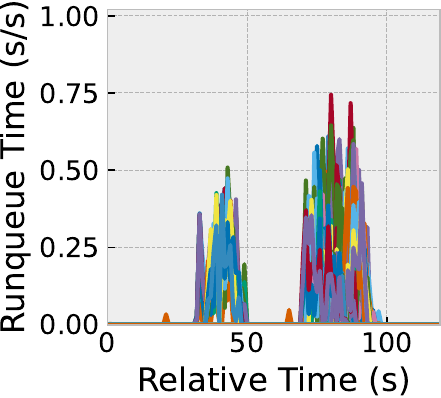}
    }\\
    \subfloat[\label{fig:ml:external_wait}]{
%    \upp\upp
      \includegraphics[width=0.3\columnwidth]{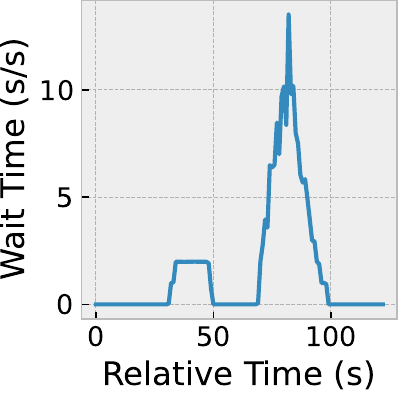}
    }
    \subfloat[\label{fig:ml:futex_wake}]{
  %  \upp\upp
      \includegraphics[width=0.32\columnwidth]{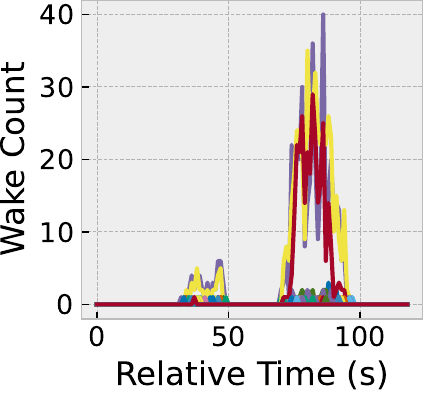}
    }\\ 
    %\upp\upp
    % \subfloat[\label{fig:ml:futex_wait}]{
    % %\upp\upp
    %   \includegraphics[width=0.25\linewidth]{graphs/ml-inference/thread_futex_wait.pdf}
    % }
    % \subfloat[\label{fig:ml:thread_sched}]{
    % %\upp\upp
    %   \includegraphics[width=0.25\linewidth]{graphs/ml-inference/thread_sched_stats.pdf}
    % } 
  %  \upp
    \cprotect\caption{\textbf{ML-inference target metric observation and performance degradation deconstruction:} (a) ML-inference load pattern; (b) ML-inference $95^{th}$ percentile response time; (c) ML-inference threads' \emph{rq\_time}; (d) ML-inference combined connection wait time; (e) Futex wake activity; } 
    \upp\upp\upp
\end{figure}

% \textbf{Deconstructing Performance Degradation:} To deconstruct performance degradation, we start by examining application's scheduling activity. Figure \ref{fig:solr:rq_time} shows server's \emph{rq\_time}, which negatively affected target metric. This suggests server's degraded performance results from CPU contention due to insufficient CPU allocation for the container.

% Despite runqueue activity coinciding with server degradation, we explore deeper connections to response time increase. \emph{Solr} uses a single epoll resource for connection multiplexing, shared by multiple waiting threads. Figure \ref{fig:solr:epoll_wait} shows these threads' total \emph{epoll\_wait\_time} for the same epoll \emph{BRI}, which dips significantly during performance degradation. Combining this with the application's connection's \emph{socket\_wait\_time} via \emph{epoll\_file\_wait} (Figure \ref{fig:solr:external_wait}), reveals \emph{rq\_time} activity impacting connection handling. All threads that waited for this epoll \emph{BRI} during degradation also showed anomalous runqueue activity in Figure \ref{fig:solr:rq_time}, suggesting the added latency from runqueue wait time impaired timely data reception and processing by connection threads.

%\paragraph{\textbf{ML-inference}}
\subsubsection{ML-inference}\label{sec:ml-inference}\textbf{Experiment Scenario \& Target Metric Observation:} As described in Section~\ref{sec:experiments}, we evaluated an ML-inference server deployed with \emph{gunicorn} configured with four worker processes. A variable load pattern (Figure~\ref{fig:ml:load}) was generated using \emph{Locust}, ranging from 2 to 60 concurrent users to induce short bursts of very high workload. The 95th percentile response time (Figure~\ref{fig:ml:target}) serves as our target metric, exhibiting a 100$\times$ increase, from \SI{75}{\milli\second} to \SI{7500}{\milli\second}, under degraded conditions.

\textbf{Deconstructing Performance Degradation:} 
We begin our analysis with the \emph{gunicorn} worker thread responsible for handling client requests. Figure~\ref{fig:ml:external_wait} shows the time this entrypoint thread spends waiting for incoming connections, closely reflecting the load pattern applied to the ML-inference server. The short, intermittent bursts of activity indicate that this worker only receives requests when load balancing assigns it new clients. Upon receiving data, the entrypoint thread delegates request handling to other threads within the same worker process via futex-based synchronization, as shown in Figure~\ref{fig:ml:futex_wake}. The resulting \emph{rq\_time} activity of these handler threads (Figure~\ref{fig:ml:rqtime}) exhibits a sharp increase in CPU contention during load surges. This behavior indicates that CPU saturation within the worker threads is the primary factor behind the ML-inference server’s degraded performance.

\begin{figure}[t]
    \centering
  %  \upp\upp\upp\upp
    \subfloat[\label{fig:redis:target}]{
      \includegraphics[width=0.35\columnwidth]{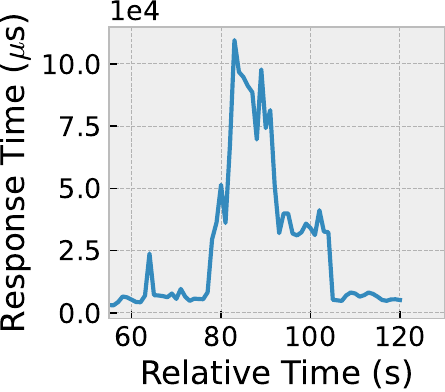}
    }
    \subfloat[\label{fig:redis:sched}]{
      \includegraphics[width=0.33\columnwidth]{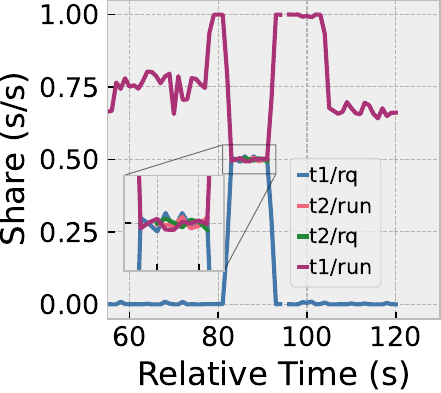}
    }
    %\vspace*{-0.2cm}
 %   \upp
    \caption{\textbf{Redis target metric observation and performance degradation deconstruction:} (a) Memtier benchmark $95^{th}$ percentile response time; (b) Redis \emph{runtime} and \emph{rq\_time} activity.} 
    \upp\upp
\end{figure}

%\paragraph{\textbf{Redis}}
\subsubsection{Redis}\label{sec:redis}\textbf{Experiment Scenario \& Target Metric Observation:} Redis experiment uses the \emph{memtier} benchmark~\cite{redis-memtier} to measure $95^{th}$ percentile latency (Figure~\ref{fig:redis:target}). As an intervention, the official Redis benchmark~\cite{redis-benchmark} runs from \SI{76}{\second} for \SI{30}{\second}, saturating the server with additional workload.

\textbf{Deconstructing Performance Degradation:} 
Given Redis’s largely single-threaded architecture, entrypoint thread (\texttt{t1}) in Figure~\ref{fig:redis:sched} is responsible for handling requests and communicating with clients. At the start of the intervention, \texttt{t1}’s runtime sharply increases. Shortly, a background Redis thread (\texttt{t2}) begins competing for same CPU core, resulting in contention that raises the observed \emph{rq\_time} for both \texttt{t1} and \texttt{t2} for approximately \SI{13}{\second}. Once \texttt{t2} terminates, \texttt{t1} resumes uninterrupted execution, and CPU usage stabilizes at $66$\%, allowing the server to efficiently handle the remaining load.

% Analysing server scheduling (Figure \ref{fig:redis:sched}) reveals two key points. Initially, the \emph{Redis} main thread's runtime increases to $100$\% at the start of the benchmark. Soon after, another Redis thread contends for the same CPU resources. This contention lasts \SI{13}{\second} until the new thread terminates. Despite the main thread regaining uninterrupted execution, server remains overwhelmed. Post-intervention, thread core usage drops to $66$\%, adequately handling the workload. Scheduling activity, combined with main thread managing all external requests via epoll, suggests CPU core contention and saturation caused server's degradation.

\subsection{Externally Constrained}
\begin{figure}[t]
    \centering
  %  \upp\upp
    \subfloat[\label{fig:teastore:load}]{
      \includegraphics[width=0.3\columnwidth]{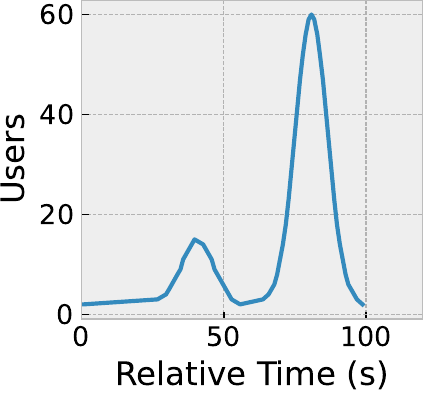}
    }
    \subfloat[\label{fig:teastore:target}]{
      \includegraphics[width=0.3\columnwidth]{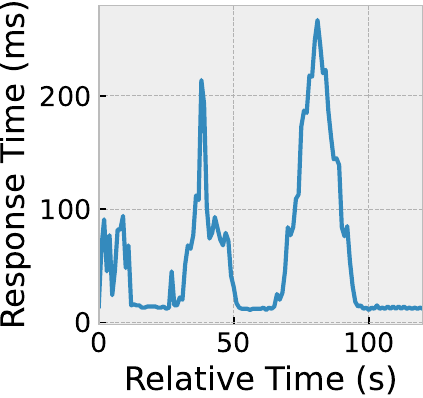}
    }
    \subfloat[\label{fig:teastore:rq_time}]{
      \includegraphics[width=0.3\columnwidth]{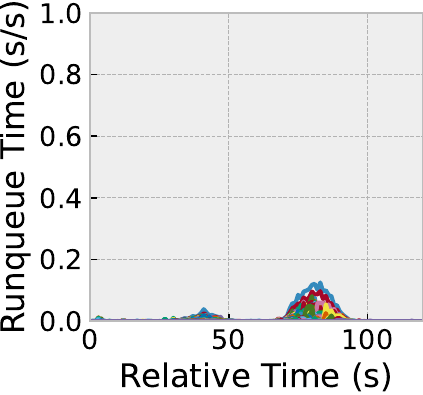}
    }\\ 
  %  \upp\upp
    \subfloat[\label{fig:teastore:runtime}]{
    \upp\upp
      \includegraphics[width=0.3\columnwidth]{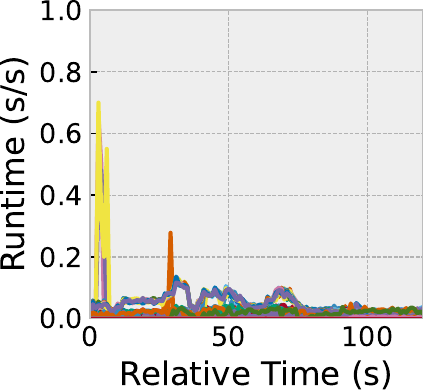}
    }
    \subfloat[\label{fig:teastore:histogram}]{
  %  \upp\upp
      \includegraphics[width=0.6\columnwidth]{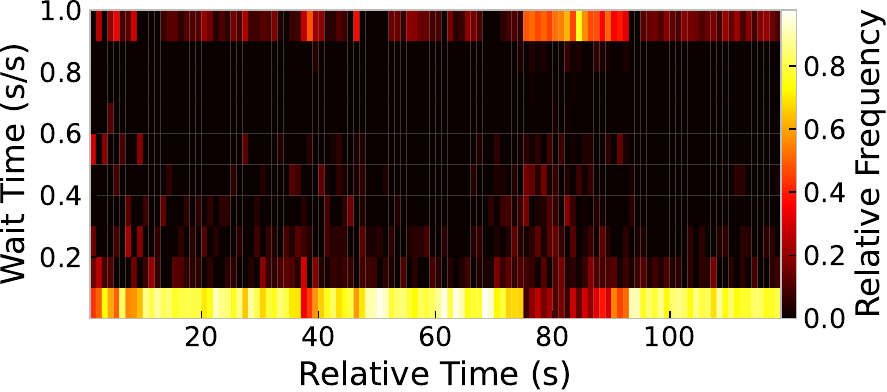}
    } 
%\upp
    \caption{\textbf{Teastore target metric observation and performance degradation deconstruction:} (a) Teastore load pattern; (b) End-to-end $95^{th}$ percentile response time; (c) Webui's threads' \emph{rq\_time}; (d) Webui's threads' \emph{runtime}; (e) Time-series histogram of webui's wait time for the constrained service.} 
   \upp\upp
\end{figure}

%\paragraph{\textbf{Teastore}}
\subsubsection{Teastore}\textbf{Experiment Scenario \& Target Metric Observation:} 
In multi-service systems, performance degradation can originate from external dependencies. Our final experiment deploys the \emph{Teastore}~\cite{von2018teastore}, monitoring the \emph{webui} component while constraining the \emph{persistence} service. A variable load pattern is applied (Figure~\ref{fig:teastore:load}), and the client’s $95^{\text{th}}$ percentile response time is measured as the target metric (Figure~\ref{fig:teastore:target}).

\textbf{Deconstructing Performance Degradation:} 
We analyze the system's \emph{rq\_time} and \emph{runtime} in Figures \ref{fig:teastore:rq_time} and \ref{fig:teastore:runtime}. The \emph{webui} component shows low CPU usage throughout, with runqueue activity increasing at peak load. This runqueue latency increase doesn't fully explain the sharp response time rise. Figure \ref{fig:teastore:histogram} presents a time series histogram of \emph{webui} service \emph{socket\_wait\_time} derived from connections to the constrained \emph{persistence} service. Higher y-axis positions indicate longer wait times, while bin brightness shows relative frequency within each timestamp column.

\begin{figure}[t]
    \centering
  %  \upp\upp\upp
    \subfloat[\label{fig:distribution:kafka-block}]{
% \upp\upp
      \includegraphics[width=0.31\columnwidth]{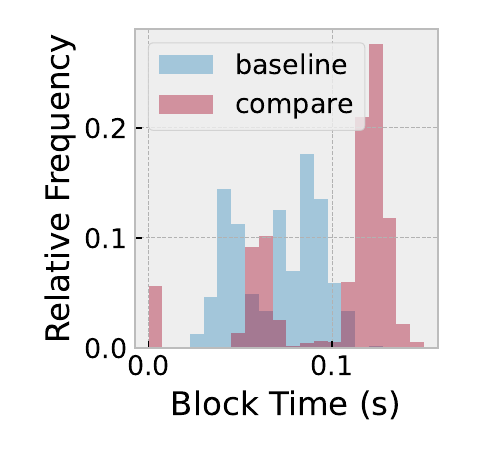}
 %     \upp\upp
    }
    \subfloat[\label{fig:distribution:kafka-pipe}]{
  % \upp\upp
      \includegraphics[width=0.31\columnwidth]{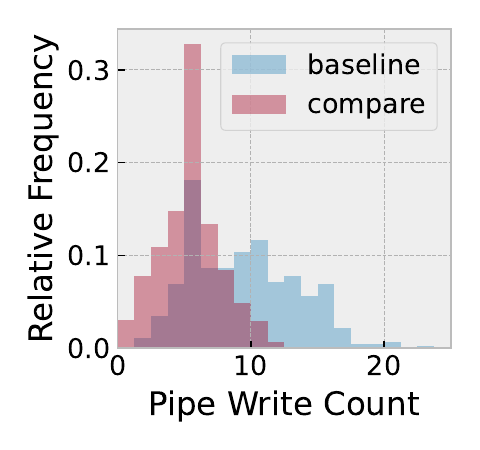} 
  %     \upp\upp
    }
    \subfloat[\label{fig:distribution:mysql-futex}]{
 %   \upp\upp
      \includegraphics[width=0.31\columnwidth]{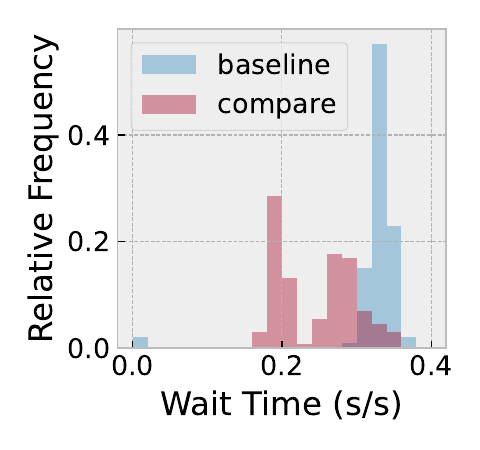}
   %    \upp\upp
    } 
    %\upp
    \caption{
        {\color{black}
        Alternative compare and baseline distribution representation for: (a) Fig~\ref{fig:kafka:block_time}; (b) Fig~\ref{fig:kafka:pipe_write}; (c) Fig~\ref{fig:mysql:disk_futex_wait}.
        } 
    }
    \label{fig:distributions}
   \upp\upp\upp\upp
\end{figure}
\noindent
\begin{minipage}{\linewidth}
%\centering
%\scriptsize
\begin{lstlisting}[
    caption={\color{black}Templated query that extracts baseline and compare distribution data for an application's thread block time.}, 
    style=sqlstyle, 
    label=lst:distribution:kafka-block,
    xleftmargin=0pt,
    xrightmargin=0pt
]
SELECT ts, blkio_share, 'baseline'
FROM taskstats_view 
WHERE {{ pid_filter }}
  AND {{ baseline_filter }}
  AND blkio_share > 0
UNION ALL
SELECT ts, blkio_share, 'compare'
FROM taskstats_view 
WHERE {{ pid_filter }}
  AND {{ compare_filter }}
  AND blkio_share > 0
\end{lstlisting}
\end{minipage}

Figure \ref{fig:teastore:histogram} highlights three key regions. The most notable, \SIrange[range-phrase=\,--\,]{70}{92}{\second} interval, shows \emph{persistence} service wait times shifting from \SIrange[range-phrase=\,--\,]{0}{0.1}{\second} to \SIrange[range-phrase=\,--\,]{0.9}{1}{\second}, coinciding with peak load (i.\,e., $60$ users). Within the \SIrange[range-phrase=\,--\,]{35}{48}{\second} interval, Figure \ref{fig:teastore:histogram} also shows an increased wait time for \emph{persistence} service connections, also coinciding with an increase in user count ($15$ users). Lastly, the \SIrange[range-phrase=\,--\,]{0}{10}{\second}
interval, aligns with the initial high latency, despite a low user count (2 users). These insights indicate the degradation is external to the \emph{webui} service, demonstrating the selected metrics' utility in identifying both internal and external sources of contention.

%-------------------------------------------------------------------------------
\section{Discussion}\label{sec:discussion}
{\color{black}
\textbf{Mitigating Metric Noise:} 
% Noisy time-series data can obscure meaningful performance signals, hindering root-cause analysis. 
To enhance diagnostic clarity, we augment time-series views with a distributional representation of metric values between baseline (i.e, normal) and comparison (i.e, degraded) periods. This shift from temporal to statistical domain can enable robust quantification of state changes using statistical tests (e.g., Mann-Whitney U, Kolmogorov-Smirnov) and effect size measures (Wasserstein distance, Cohen’s d).

For example, Figure~\ref{fig:distribution:kafka-block} transforms the noisy \emph{block\_time} signal from  Figure~\ref{fig:kafka:block_time} into two distinct distributions, clearly revealing a shift toward higher thread block wait times. Similarly, Figure~\ref{fig:distribution:kafka-pipe} shows reduced pipe write frequency in Kafka’s degraded phase from Figure~\ref{fig:kafka:pipe_write}, while Figure~\ref{fig:distribution:mysql-futex} highlights decreased futex wait times in MySQL from Figure~\ref{fig:mysql:disk_futex_wait}, which is consistent with increased thread activity in other subsystems. 
These distributional views are generated through templated queries integrated into our diagnostic workflow, allowing systematic, statistically grounded comparison without manual inspection of volatile time-series data. For example, Listing~\ref{lst:distribution:kafka-block} extracts an application’s block-time distribution using parameterized user input.

}

% \begin{wraptable}[9]{R}{0.45\columnwidth}
% \centering
% \caption{\textbf{Instrumentation overhead}: Latency mean and stdev with(out) instrumentation.}
% %\upp
% \label{table:overhead}
% \setlength{\tabcolsep}{0.5pt}
% \resizebox{0.5\columnwidth}{!}
% {
% \begin{tabular}{|c|c|c|c|}
% \hline
% \textbf{App} & 
% \multicolumn{1}{c|}{\textbf{\begin{tabular}[c]{@{}c@{}}Latency\textsubscript{w/out}\\ Mean ± SD (ms)\end{tabular}}} & 
% \multicolumn{1}{c|}{\textbf{\begin{tabular}[c]{@{}c@{}}Latency\textsubscript{with}\\ Mean ± SD (ms)\end{tabular}}} & 
% \multicolumn{1}{c|}{\textbf{\begin{tabular}[c]{@{}c@{}}Latency \\ Overhead (ms)\end{tabular}}} \\
% \hline
% Redis       &  $2.053 \pm 0.155$ & $2.425 \pm 0.310$ & $0.372$ \\ \hline
% Cassandra   &  $1.689 \pm 0.002$ & $1.758 \pm 0.007$ & $0.070$ \\ \hline
% MySQL       & $47.193 \pm 2.344$ & $47.495 \pm 3.023$ & $0.302$ \\ \hline
% Kafka       &  $2.170 \pm 0.051$ & $2.231 \pm 0.033$ & $0.061$ \\ \hline
% \end{tabular} 
% }

% \end{wraptable}
\textbf{Overhead Analysis:} We evaluate overhead for four standalone applications using the same workloads from Section~\ref{sec:experiments}, comparing latency with(out) our instrumentation over $10$ runs. Table~\ref{table:overhead} shows the latency mean and standard deviation for both scenarios. Induced instrumentation overhead depends on the frequency and computation of our \emph{eBPF} programs and varies with each application's kernel resource needs. We observe that instrumentation impacts Cassandra and Kafka least ($\approx$ \SI{0.07}{\milli\second}) and Redis and MySQL most ($\approx$ \SI{0.3}{\milli\second}), likely because MySQL and Redis process requests immediately, while Kafka and Cassandra defer processing. 
{\color{black}
Moreover, under saturation benchmarks (i.e., presented in Sections~\ref{sec:kafka}, \ref{sec:ml-inference}, \ref{sec:redis}), where Kafka, Redis, and ML‑inference were heavily loaded, our tool’s overhead remained low: i.e., 10\% of a single CPU core and ~115KB/s disk writes. This efficiency is attributable to eBPF’s in‑kernel statistical aggregation, which minimizes userspace processing.
% Additionally, during benchmarks that saturated the application and system (Sections~\ref{sec:kafka}, \ref{sec:redis} and \ref{sec:ml-inference}), our tool's CPU usage and disk write rate were 10\% of a single core and ~115KB/s, a low overhead that is attributable to eBPF’s in-kernel statistical aggregation design, which minimizes userspace processing.
}

\begin{table}
%\upp
\centering
\caption{\textbf{Instrumentation overhead}: Latency mean and stdev with(out) instrumentation.}
%\upp
\label{table:overhead}
\setlength{\tabcolsep}{0.5pt}
\resizebox{0.9\columnwidth}{!}
{
\begin{tabular}{|c|c|c|c|}
\hline
\textbf{App} & 
\multicolumn{1}{c|}{\textbf{\begin{tabular}[c]{@{}c@{}}Latency\textsubscript{w/out}\\ Mean ± SD (ms)\end{tabular}}} & 
\multicolumn{1}{c|}{\textbf{\begin{tabular}[c]{@{}c@{}}Latency\textsubscript{with}\\ Mean ± SD (ms)\end{tabular}}} & 
\multicolumn{1}{c|}{\textbf{\begin{tabular}[c]{@{}c@{}}Latency \\ Overhead (ms)\end{tabular}}} \\
\hline
Redis       &  $2.053 \pm 0.155$ & $2.425 \pm 0.310$ & $0.372$ \\ \hline
Cassandra   &  $1.689 \pm 0.002$ & $1.758 \pm 0.007$ & $0.070$ \\ \hline
MySQL       & $47.193 \pm 2.344$ & $47.495 \pm 3.023$ & $0.302$ \\ \hline
Kafka       &  $2.170 \pm 0.051$ & $2.231 \pm 0.033$ & $0.061$ \\ \hline
\end{tabular} 
} 
\upp\upp
\end{table}

\section{Conclusion}\label{sec:conclusion}
Application performance degradation diagnosis requires overcoming the constraints of thread state analysis to uncover the thread dynamics that underpin performance bottlenecks. In this paper, we presented an application-agnostic performance degradation analysis method that extends TSA by leveraging \emph{eBPF} to instrument fine-grained thread-resource interactions across six critical kernel subsystems. The $16$ implemented metrics captures the precise relationships between threads and resources like futexes, sockets, and disks, further revealing how degradation propagates through an application's internal architecture. 
{\color{black}
While our current focus is on thread–resource interactions, extending this metric set with complementary system-wide statistics is planned for future work.
}
We demonstrated the efficacy of our approach by validating it through extensive experiments on a diverse set of six OLDI applications, including MySQL, Kafka, and Redis. Under various workload and contention scenarios, the collected metrics successfully pinpointed the root causes of degradation, such as identifying lock-contending threads, disk IO bottlenecks, and CPU saturation. Furthermore, our method showed capability of distinguishing internal bottlenecks from external service dependencies (e.g., TeaStore microservice benchmark). More importantly,  this granular observability was achieved by our method with minimal instrumentation overhead, confirming its feasibility for real-world deployment.

\section*{Acknowledgment}
%\small
This work was supported and funded by European Union (EU) \emph{CETPartnership} (CET-P) FLEXI project and \emph{Dutch Research Council} (NWO) (Grant EP.1602.24.003).

\bibliographystyle{IEEEtran}
\bibliography{arxiv}

\input{ad-ae}

\end{document}

%% file: ad-ae.tex
% Uncomment/comment the following usetag lines 
% if you would like to get an explanation or an example.
% Don't forget to comment them for the final submission.
% \usetag{explanation}
% \usetag{example}

\twocolumn[%
{\begin{center}
\Huge
\textbf{Appendix: Artifact Description/Artifact Evaluation}  
\vspace{0.6cm}
\end{center}}
]
\setcounter{footnote}{0}
\setcounter{figure}{0}

This two-page appendix contains the \emph{Artifact Description (AD)} and \emph{Artifact Evaluation (AE)}.  The complete source code for the metric collector and analysis interface is also released opensource in an \emph{online} \verb|GitHub| repository~\cite{repo}. The experiment datasets, the notebooks used for analysis, and detailed instructions/scripts to reproduce the experiments and results presented in Sections~\ref{sec:experiments} and~\ref{sec:results} are publicly available at \url{https://github.com/EC-labs/ipdps2026-prism-artifact/releases/download/v1.0.0/archive.tar}.

%%%%%%%%%%%%%%%%%%%%%%%%%%%%%%%%%%%%%%%%%%%%%%%%%%%%%
%  AD Appendix
%%%%%%%%%%%%%%%%%%%%%%%%%%%%%%%%%%%%%%%%%%%%%%%%%%%%%

\appendixAD \label{sec:ad}
 \setcounter{section}{0}

\section{Overview of Contributions and Artifacts}

\subsection{Paper's Main Contributions}

\begin{itemize}[label={}]
    \item $C_!$: A comprehensive set of metrics that capture an application's thread dynamics;
    \item $C_2$: A metric collector (Prism) that gathers the metrics listed in Table~\ref{table:metrics} and stores them in a structured database for analysis;
    \item $C_3$: A user interface with templated queries that enable exploration of an application's thread dynamics through the metrics database.
\end{itemize}

\subsection{Computational Artifacts}

\begin{itemize}[label={}]
    \item $A_1$: \url{https://github.com/EC-labs/ipdps2026-prism-artifact/releases/download/v1.0.0/archive.tar}
\end{itemize}

\begin{center}
\begin{tabular}{rll}
\toprule
Artifact ID  &  Contributions &  Related \\
             &  Supported     &  Paper Elements \\
\midrule
$A_1$   &  $C_1$, $C_2$, $C_3$ & Figures 6-14 \\
\bottomrule
\end{tabular}
\end{center}

%%%%%%%%%%%%%%%%%%%%%%%%%%%%%%%%%%%%%%%%%%%%%%%%%%%%%%%%
\section{Artifact Identification}
%%%%%%%%%%%%%%%%%%%%%%%%%%%%%%%%%%%%%%%%%%%%%%%%%%%%%%%%

\newartifact

\artrel

This artifact $A_1$ provides the instructions and data necessary to reproduce the analysis in Sections~\ref{sec:experiments} and \ref{sec:results} pertaining to $C_1$-$C_3$.

\artexp
Assuming similar computational resources to those indicated in Section~\ref{sec:experiments}, the application should illustrate similar thread dynamics and degradation patterns to those presented in Section~\ref{sec:results}.

\arttime

Without analysis, reproducing the $6$ experiments in Section~\ref{sec:experiments} requires approximately \SI{30}{\minute}: \SI{1}{\minute} to setup the target application and metric collector (Prism), \SI{4}{\minute} to execute the workloads and degradation scenarios, and another \SI{1}{\minute} for post-processing the target KPI metrics to the format expected by Prism's analysis UI. 

With regard to the time for analysis, this may vary between experiments assuming the Selective Thread Tracking Algorithm~\ref{algo:selective_tracking} is followed. Nevertheless, to facilitate this process, Prism provides an interface with which one can navigate the collected metrics, and templated queries.

\artin

\artinpart{Hardware}

The artifact has no strict hardware requirements, although adequate computational resources are recommended for replication purposes. Our setup, however, consisted of a \verb|i7-11850H Intel| server @ \SI{2.50}{\giga\hertz} running  Ubuntu 22.04 LTS, Jammy Jellyfish (\verb|x86_64|) OS with \verb|6.8.12| Linux kernel, $8$ CPU cores, $2$$\times$\SI{16}{\giga\byte} \verb|DDR4| RAM, and \SI{1}{\tera\byte} SSD disk. 

\artinpart{Software}

The artifact was evaluated on a machine with the Linux \verb|6.8.12| kernel (with BPF/BTF enabled). Additionally, the artifact also requires installing Docker and the Nix package manager. To fully reproduce our configuration, we installed Docker \verb|27.5.1| and Nix \verb|2.28.4|. The remaining software dependencies, and their respective versions are pinned by the provided \texttt{flake.nix} and \texttt{flake.lock} files in the root directory of the artifact.

\artinpart{Datasets / Inputs}

The artifact's \verb|flake.nix| and \verb|flake.lock| files automatically manage the \emph{dataset} and \emph{input} dependencies for the experiments. As such, these are downloaded and resolved when an experiment's run script is executed.
% Nevertheless, the input dependencies to reproduce the experiments are: \emph{Redis}, \emph{Redis Bench} \emph{Memtier Benchmark}

\artinpart{Installation and Deployment}

Some experiments (e.g., \emph{Cassandra} and \emph{Kafka}) require a running Docker daemon. Similar to the previous point, the remaining \emph{installation} and \emph{deployment} dependencies are managed and pinned by the \verb|flake.*| files in the artifact's root directory.

\artexpl{
    Detail the requirements for compiling, deploying, and executing the experiments, including necessary compilers and their versions.
}

\artcomp

Each of the six experiments includes the following stages: \textbf{(1) Setup}: Downloading required Docker images and deploying the target application. \textbf{(2) Execution}: Starting and bootstrapping Prism's metric collector on the application, and managing workloads and degradation scenarios. \textbf{(3) Post-process}: Raw KPI metrics (e.g., per-request latency) may be post-processed into more informative metrics (e.g., 95th percentile latency). \textbf{(4) Cleanup}: Experiment services and resources are terminated. \textbf{(5) Analysis}: The metric collector's data and target KPI metric are uploaded to Prism's UI for analysis.

% Each of the six experiments contain the following stages: \textbf{(1) Setup}: This usually consists of downloading the required docker images, and deploying the application that is to be analyzed (e.g. Kafka); \textbf{(2) Execution}: Prism's metric collector starts and is bootstrapped on the target application. This stage also manages the experiment workloads and degradation scenarios. \textbf{(3) Post-process}: The raw KPI metrics (e.g. per-request latency) collected throughout the execution stage may be post-processed into more informative metrics (e.g. 95th percentile latency). \textbf{(4) Cleanup}: The services and resources created for the experiment are terminated. \textbf{(5) Analysis}: Via Prism's analysis interface, the data extracted by Prism's metric collector and the target KPI metric are uploaded to the UI for analysis. 

\artexpl{
    Provide an abstract description of the experiment workflow of the artifact. It is important to identify the main tasks (processes) and how they depend on each other. 
    
    A workflow may consist of three tasks: $T_1, T_2$, and $T_3$. The task $T_1$ may generate a specific dataset. This dataset is then used as input by a computational task $T_2$, and the output of $T_2$ is processed by another task $T_3$, which produces the final results (e.g., plots, tables, etc.). State the individual tasks $T_i$ and provide their dependencies, e.g., $T_1 \rightarrow T_2 \rightarrow T_3$.
    
    Provide details on the experimental parameters. How and why were parameters set to a specific value (if relevant for the reproduction of an artifact), e.g., size of dataset, number of data points, input sizes, etc. Additionally, include details on statistical parameters, like the number of repetitions.
}

\artout

Analysis requires uploading the \verb|*.db3| metric database and target KPI metric files to Prism's UI, which then provides pre-existing analyses like a process-level dependency map showing processes interacting with the bootstrapping process. A \verb|Debug| page allows executing any query against the backing \verb|*.db3| database. To facilitate Selective Thread Tracking, we provide templated queries in the UI's \href{https://github.com/EC-labs/prism/tree/master/analysis/src/sql}{analysis/src/sql} directory. Each query can generate line, histogram, or bar plots if the resulting data supports the plot type.

% The analysis requires uploading the \verb|*.db3| metric database and the target KPI metric files to Prism's analysis UI. After doing so, the UI provides some pre-existing analysis such as the process-level dependency map, which indicates the processes that directly and indirectly interacted with the bootstrapping process. There is also a \verb|Debug| page that allows executing any query against the backing \verb|*.db3| database. Nevertheless, to facilitate the Selective Thread Tracking process, we provide templated queries in the UI's \href{https://github.com/EC-labs/prism/tree/master/analysis/src/sql}{analysis/src/sql} directory. Each query can then be plotted into their respective line, histogram, or bar plots, as long as the data resulting from the query supports the desired plot type.

%%%%%%%%%%%%%%%%%%%%%%%%%%%%%%%%%%%%%%%%%%%%%%%%%%%%%
%  AE Appendix
%%%%%%%%%%%%%%%%%%%%%%%%%%%%%%%%%%%%%%%%%%%%%%%%%%%%%
%\newpage
\appendixAE
\setcounter{section}{3}

% The following artifact evaluation description leverages the \emph{Redis} experiment for ease of explainability, however, the evaluation process for the remaining experiments is similar to that of \emph{Redis}. We also assume that the software dependencies mentioned in the \emph{Artifact Description} (i.e. Nix, Docker, and a sufficiently recent version of Linux with eBPF enabled) are already available.

The artifact evaluation uses the \emph{Redis} experiment as an example; the process for other experiments is similar. We assume the dependencies listed in the \emph{Artifact Description} (i.e., Nix, Docker, and a recent Linux version with eBPF enabled) are already installed.

\arteval{1}
\artin

Start by downloading and unpacking the provided artifact, and changing into the unpacked archive's directory:
\begin{lstlisting}[style=terminal]
$ mkdir archive && \
    tar -xvf archive.tar -C archive && \
    cd archive
\end{lstlisting}

The archive root contains folders for each experiment in Section~\ref{sec:experiments}, each with its dependency management and run scripts. It also includes \verb|flake.nix| and \verb|flake.lock|, which pin the software dependencies.

% In the archive's root, there is a folder for each of the experiments in Section~\ref{sec:experiments}. These folders include the software dependency management and scripts required to run each experiment. The root directory also includes a \verb|flake.nix| and a \verb|flake.lock|. These files are critical to pin the software dependencies for each experiment.

\artexpl{
Provide instructions for installing and compiling libraries and code. 
Offer guidelines on deploying the code to resources.
}

\artcomp

% Running the experiment is done using the \verb|nix run| command, pointing to the flake at the root of the repository, and selecting one of the experiments, and passing a directory to the execution script where the results should be stored. For example, the following command would run the \emph{Redis} experiment, and store the results in the \verb|./results| directory:

Run an experiment with \verb|nix run|, using the flake in the repository root, select an experiment, and pass a directory for storing results. For example, the following command runs \emph{Redis'} experiment and stores results in \verb|./results|:

\begin{lstlisting}[style=terminal]
$ NIX_CONFIG='experimental-features = nix-command flakes' nix run .#redis-experiment ./results
\end{lstlisting}

Root privileges are required to start and terminate Prism's eBPF collector, so the user is prompted for the root password after setup.

% Note that to start and terminate Prism's metric collector, root privileges are required to inject eBPF programs. As such, after the setup stage, the user is prompted for the root password to be able to start the collector.

The aforementioned command fully executes the Redis experiment. The following description explains the script's dependencies and activities; \textbf{the listed commands are for documentation only and should not be executed}.

The software dependencies for \emph{Redis'} execution script are listed in \verb|redis/default.nix|. A subset is shown below:
% The aforementioned command would fully execute the Redis experiment, nevertheless, the following description serves as an explain the experiment's dependencies and the contents of the execution script. As such, \textbf{the following listed commands aren't to be executed} and are listed merely to document the script's activities.

% The software dependencies for \emph{Redis'} execution script are listed in the archive's \verb|redis/default.nix| file. The following listing introduces a subset of the dependencies:
\begin{lstlisting}[style=terminal]
runtimeDeps = with pkgs; [
    bash
    docker
    prism-mc
    packages.redis
    packages.memtier
    # ...
];
\end{lstlisting}

The dependencies include standard packages (e.g., \verb|docker|, \verb|bash|), Prism's metric collector (\verb|prism-mc|), and workload packages (\verb|packages.{redis,memtier}|). The \verb|nix run| command ensures these are available at runtime. The Redis execution script at \verb|redis/run.sh| begins with the setup stage, downloading the Redis image and starting a container.

% The dependencies, list a set of standard packages (e.g. \verb|docker| and \verb|bash|), Prism's metric collector (\verb|prism-mc|), and packages that setup the workloads for the experiment (\verb|packages.{redis,memtier}|). Before executing the script, the \verb|nix run| command ensures all dependencies are adequately available at runtime to the execution script.

% The Redis execution script can be found in the archive's \verb|redis/run.sh| file. The script starts with the setup stage which sets up the environment to run the experiment. As such, it downloads the redis image and starts a container:
\begin{lstlisting}[style=terminal]
$ docker pull redis:7.4.2
$ docker run # ... # redis:7.4.2
\end{lstlisting}

% The execution stage bootstraps the Prism metric collector on the processes running redis, and manages the instantiation of the workload generators that stress the redis server:
The execution stage bootstraps the Prism metric collector on the Redis processes and manages starting the workload generators to stress the server:
\begin{lstlisting}[style=terminal]
$ metric-collector # ... --pids <pid-list>
$ memtier_benchmark # ... --test-time 120
$ redis-benchmark # ...
\end{lstlisting}

The post-processing stage computes the $95^{th}$ percentile latency from the per-request latency generated by the memtier benchmark:
\begin{lstlisting}[style=terminal]
$ percentile-ps # ... --percentile 0.95
\end{lstlisting}

Finally, on script exit, the cleanup function is executed to terminate the resources created throughout the experiment, i.e., the redis server and Prism's metric collector:
\begin{lstlisting}[style=terminal]
$ sudo kill -SIGINT "$PRISM_PID"
$ docker stop redis
\end{lstlisting}

\artout

To analyse the data, start the analysis interface:
\begin{lstlisting}[style=terminal]
$ NIX_CONFIG='experimental-features = nix-command flakes' nix run github:ec-labs/prism#analysis 
\end{lstlisting}

\begin{figure}[h!]
    \centering
    \upp\upp\upp
    \subfloat[\label{fig:KPI}]{
      \includegraphics[width=0.33\columnwidth]{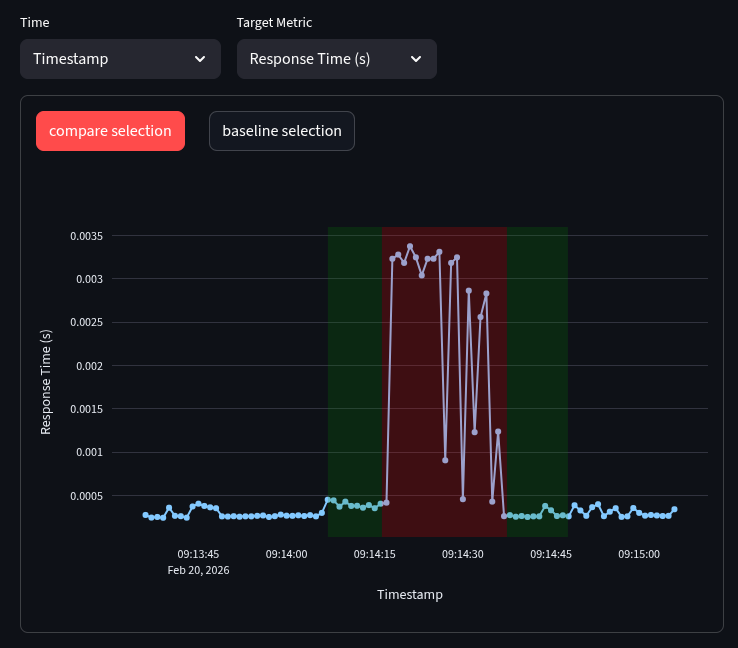}
    }
    \subfloat[\label{fig:Ripple}]{
      \includegraphics[width=0.26\columnwidth]{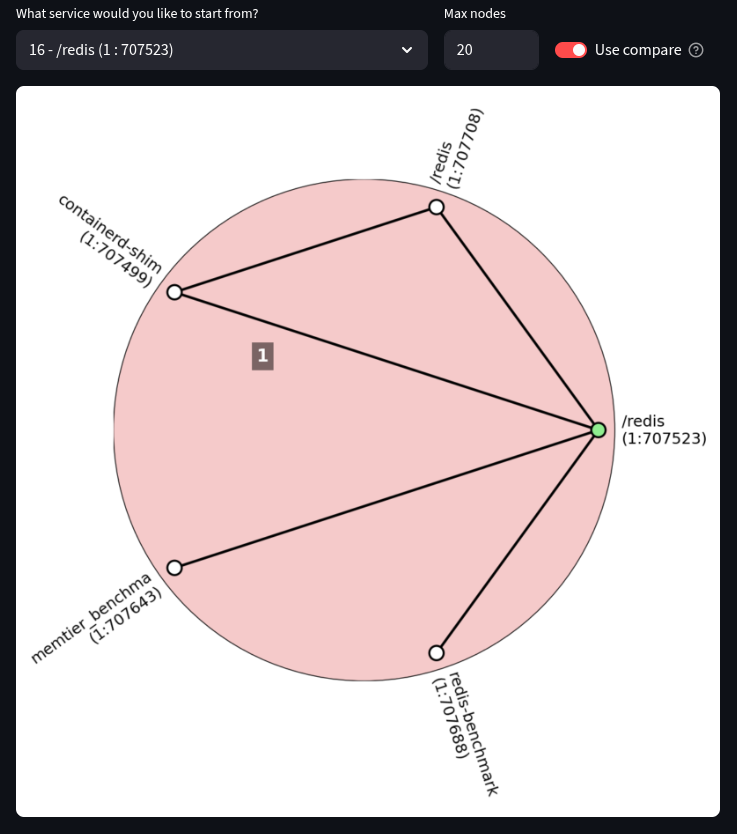}
    }
    \subfloat[\label{fig:Debug}]{
      \includegraphics[width=0.33\columnwidth]{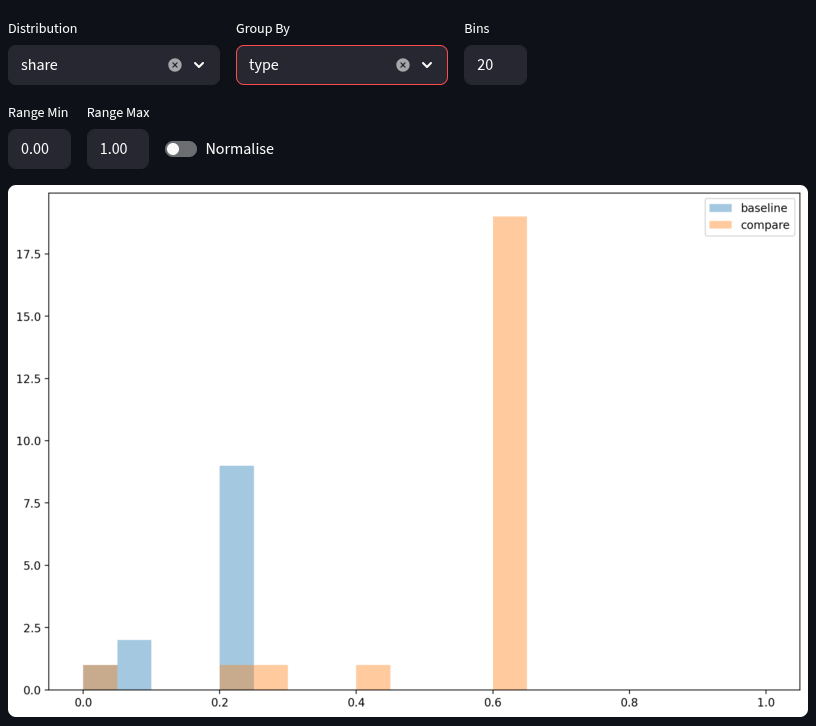}
    }
    \caption{\textbf{User interface screenshots}: (a) KPI page; (b) Ripple page; (c) Debug page.}
\end{figure}

On the interface's homepage \url{http://localhost:8501}, upload the \verb|*.db3| database file from the \verb|./results/<experiment>| directory. Then, on the \href{http://localhost:8501/kpi}{KPI} page, upload the target metric file. For \emph{Redis} experiment, selecting baseline and comparing periods displays a graph with highlighted periods, as in Figure~\ref{fig:KPI}.

% On the interface's homepage \url{http://localhost:8501}, upload the \verb|*.db3| database file generated when running the experiment. It should be under the \verb|./results| directory passed as an argument to the execution command. After doing so, you may also upload the target metric file in the \href{http://localhost:8501/kpi}{KPI} page. Using the \emph{Redis} experiment as an example, after selecting the baseline and compare periods, the KPI page displays a graph with the compare and baseline periods highlighted, as illustrated in Figure~\ref{fig:KPI}.

The baseline and compare periods serve as template variables. The \href{http://localhost:8501/ripple}{Ripple} page shows a process-level dependency graph. Specifying \emph{redis} as the bootstrapping process yields a graph like Figure~\ref{fig:Ripple}.

Lastly, the Debug page allows executing custom/templated queries. Running the \href{https://github.com/EC-labs/prism/blob/master/analysis/src/sql/distributions/pid_rq_share.sql}{pid runqueue share} query with redis' pid as \verb|pid_filter| (e.g., \verb|(pid=707523)|) and selecting histogram plot yields a plot like Figure~\ref{fig:Debug}.

% The baseline and compare periods are used as template variables in the templated queries. The  \href{http://localhost:8501/ripple}{Ripple} page provides a process-level overview of the dependencies stemming from a bootstrapping process. Upon specifying redis as the bootstrapping process, the page displays the process-level dependency graph similar to that of Figure~\ref{fig:Ripple}.

% Lastly, the Debug page allows executing custom/templated queries against the metric database. Running the \href{https://github.com/EC-labs/prism/blob/master/analysis/src/sql/distributions/pid_rq_share.sql}{pid runqueue share} query, placing redis' pid as a \verb|pid_filter| template variable (e.g. \verb|(pid=707523)|), and selecting the histogram plot, provides a plot similar to that shown in Figure~\ref{fig:Debug}.

% \artexpl{
% \begin{itemize}
%     \item Provide a description of the expected results and a methodology for evaluating these results. 
%     \item Explain how the expected results from the experiment workflow correlate with the contributions stated in the article. 
%     \item For example, if the article presents results in a figure, the artifact evaluation should also produce a similar figure, depicting the same generalizable outcome. Authors must focus on these aspects to reduce the time required for others to understand and verify an artifact.
% \end{itemize}
% }